\documentclass[epj]{svjour}
 \usepackage{lineno} 
 \usepackage{color}
 \definecolor{MPPgreen}{RGB}{0,128,112}
\usepackage{graphicx}
\usepackage{dcolumn}
\usepackage{bm}
\usepackage{amssymb}

\begin{document}
\bibliographystyle{spphys} 

\title{The {\huge$\nu$}-cleus experiment: A gram-scale fiducial-volume cryogenic detector for the first detection of coherent neutrino-nucleus  scattering}


\author{R. Strauss\inst{1}\thanks{Corresponding author: strauss@mpp.mpg.de} \and J. Rothe\inst{1} \and G. Angloher\inst{1} \and A. Bento\inst{2} \and A. G\"utlein\inst{3} \and D. Hauff\inst{1} \and H. Kluck\inst{3} \and M. Mancuso\inst{1} \and L. Oberauer\inst{4} \and F. Petricca\inst{1} \and F. Pr\"obst\inst{1} \and J. Schieck\inst{3} \and S. Sch\"onert\inst{4} \and W. Seidel\inst{1}\thanks{Deceased 19 February 2017} \and L. Stodolsky\inst{1}}


\institute{Max-Planck-Institut f\"ur Physik,  D-80805 M\"unchen, Germany \and CIUC, Departamento de Fisica, Universidade de Coimbra, P3004 516 Coimbra, Portugal \and Institut f\"ur Hochenergiephysik der \"Osterreichischen Akademie der Wissenschaften, A-1050 Wien, Austria,\\ and Atominstitut, Vienna University of Technology, A-1020 Wien, Austria \and Physik-Department, Technische Universit\"at M\"unchen, D-85748 Garching, Germany}

\date{\today}

\abstract{
We discuss a small-scale experiment, called {\large$\nu$}-cleus, for the first detection of coherent neutrino-nucleus scattering
by probing nuclear-recoil energies down to the 10\,eV-regime. The  detector consists of low-threshold CaWO$_4$ and  Al$_2$O$_3$  calorimeter arrays with a total mass of about 10\,g and several cryogenic veto detectors operated at millikelvin temperatures. Realizing a fiducial volume and a multi-element target, the detector enables active discrimination of $\gamma$, neutron and surface   backgrounds.  A first prototype Al$_2$O$_3$ device, operated above ground in a setup without shielding, has achieved an energy threshold of ${\sim20}$\,eV  and further improvements are in reach.  A sensitivity study for the detection of coherent neutrino scattering at nuclear power plants shows a unique discovery potential (5$\sigma$) within a measuring time of ${\lesssim2}$ weeks.  Furthermore, a site at a thermal research reactor and the use of a radioactive neutrino source are investigated. With this technology, real-time monitoring of nuclear power plants  is feasible.
}

\titlerunning{The {\large$\nu$}-cleus experiment: A gram-scale fiducial-volume cryogenic detector\dots}

\PACS{
{5.55.Vj}{Neutrino, muon, pion, and other elementary particle detectors; cosmic ray detectors }
}

\maketitle

\section{Introduction}
\label{sec:intro}
The detection of coherent neutrino-nucleus scattering \linebreak (CNNS) is among the most challenging tasks of modern particle and astroparticle physics.  A first observation of CNNS would be an important confirmation of the Standard Model of Particles and would open the door to new physics beyond the Standard Model (BSM). 

Coherent neutrino-nucleus scattering (CNNS), first proposed in 1974 \cite{PhysRevD.9.1389},  is an unobserved neutral-current interaction predicted by the Standard Model of Particle Physics. Neutrino-nucleus scattering via Z$_0$-exchange becomes coherent over the nuclei at low transferred momenta, for large nuclei simultaneously boosting the interaction cross-section and reducing the recoil energies. 
The total elastic cross section for the process can be written as \cite{PhysRevD.30.2295}:
\begin{small}

\begin{eqnarray}\label{equ:sigma}
\frac{d\sigma}{dE_R}& =& \frac{G_F^2}{8\pi(\hbar c)^4} \left((4 \sin^2\theta_W-1)\cdot Z+N\right)^2 \nonumber\\
&& \cdot m_N \cdot (2 -E_R m_N /E_\nu^2)|f(q)|^2
\end{eqnarray}
\end{small}
where $G_F$ is Fermi's coupling constant, $\theta_W$ the Weinberg angle,  $Z$, $N$ and $m_N$ are the nucleus' 
proton number, neutron number, and total mass respectively, $E_\nu$ is the neutrino  energy, and $E_R$ the 
resulting nuclear-recoil energy.    The nuclear form factor $f(q)$ describes the loss of coherence as a 
function of transferred momentum wavenumber $q=\sqrt{2 m_N E_R}/\hbar$. It can be understood as the Fourier 
transform of the nuclear weak charge density, and is close to unity for small $q$ (typically at 
$E_\nu\lesssim50$\,MeV).

The process remains unobserved until now due to the small recoil energies expected which challenge detector technologies. Multiple experimental efforts  for detecting CNNS are taken globally: the COHERENT experiment~\cite{Akimov:2015nza} which is currently taking data at the Spallation Neutron  Source (SNS) uses a combination of conventional CsI, Ge and liquid-Xe detectors. Various other experiments are planned or being commissioned such as CONNIE~\cite{Aguilar-Arevalo:2016khx} using CCDs~\cite{Moroni:2014wia}, TEXONO~\cite{Kerman:2016jqp} using ionization-based Ge detectors,  and MINER~\cite{2016arXiv160902066M} and RICOCHET~\cite{Billard:2016giu} using cryogenic detector technology.   

The ultra-low-threshold cryogenic calorimeters present\-ed  here (and in \cite{cnns_letter}) put a rapid detection of this process within reach technologically. The relatively large cross section compared e.g. to neutrino-electron scattering makes this experimental approach interesting in two ways: 1) CNNS is detectable with a small-scale experiment  and a total target mass of 10\,g within a measuring time of several weeks (see below), far less costly than traditional neutrino facilities. 2) A managable scaling of the total target mass to the still moderate range of 1-10\,kg opens up a new window for precision tests of neutrino properties and interactions beyond the standard model. A recent summary of CNNS sensitivity to BSM neutrino physics is given in~\cite{Lindner:2016wff}, including the following potential observations.



\begin{itemize}
 \item Interpreted within the standard model, a precise measurement of the CNNS cross-section allows to determine the Weinberg angle at low energy scale through Equ.~\ref{equ:sigma}. Transferred momenta in CNNS are on the order of few $\mathrm{MeV}/\mathrm{c^2}$, extending the reach of other planned low momentum-transfer precision experiments \cite{2015arXiv151103934B}. Together with knowledge on electroweak precision observables (e.g. from LEP), this allows to probe the running of the weak mixing angle~\cite{Lindner:2016wff} which is precisely predicted in the standard model~\cite{PhysRevD.72.073003}. This collective measurement has sensitivity to BSM contributions well above the LHC scale.
 \item The neutrino-quark sector of neutrino Non-Standard Interactions~\cite{0034-4885-76-4-044201,santamaria2003present}, i.e. modified V-A quark-neutrino couplings, may measurably modify the CNNS cross-section~\cite{Lindner:2016wff}.
 \item Exotic Neutral Currents~\cite{Lindner:2016wff}, i.e. general \mbox{(pseudo-)}scalar, \mbox{(axial-)}vector or tensor couplings can induce modifications in the CNNS cross section and energy spectrum.
 \item The possibility of observing active-to-sterile neutrino oscillations using CNNS is discussed in~\cite{PhysRevD.86.013004}.
 \item For very low energy thresholds, the magnetic moment of the neutrino (causing enhanced low-energy scattering with spin exchange) can be probed beyond current limits from neutrino-electron scattering~\cite{doi:10.1142/S0217732305017482}.
 
\end{itemize}


\section{The  detector}

\subsection{A fiducial-volume cryogenic detector}
A detector, sensitive to CNNS, faces two main challenges: an extremely low energy threshold combined with extraordinarily small background levels. We present a new  gram-scale cryogenic detector which combines the possibility of lowest nuclear-recoil thresholds ($\mathcal{O}$(${\lesssim}10$\,eV)) and the advantages of a fiducial volume device.   Those provide active shielding  by the outermost regions against external radiation which reduces  the background level in the innermost target volume (the fiducial volume).  Since an exact spatial position reconstruction of events is difficult to realize in thermal detectors, so far this potential could not be exploited. 

\begin{figure}
\centering
\includegraphics[width=0.35\textwidth]{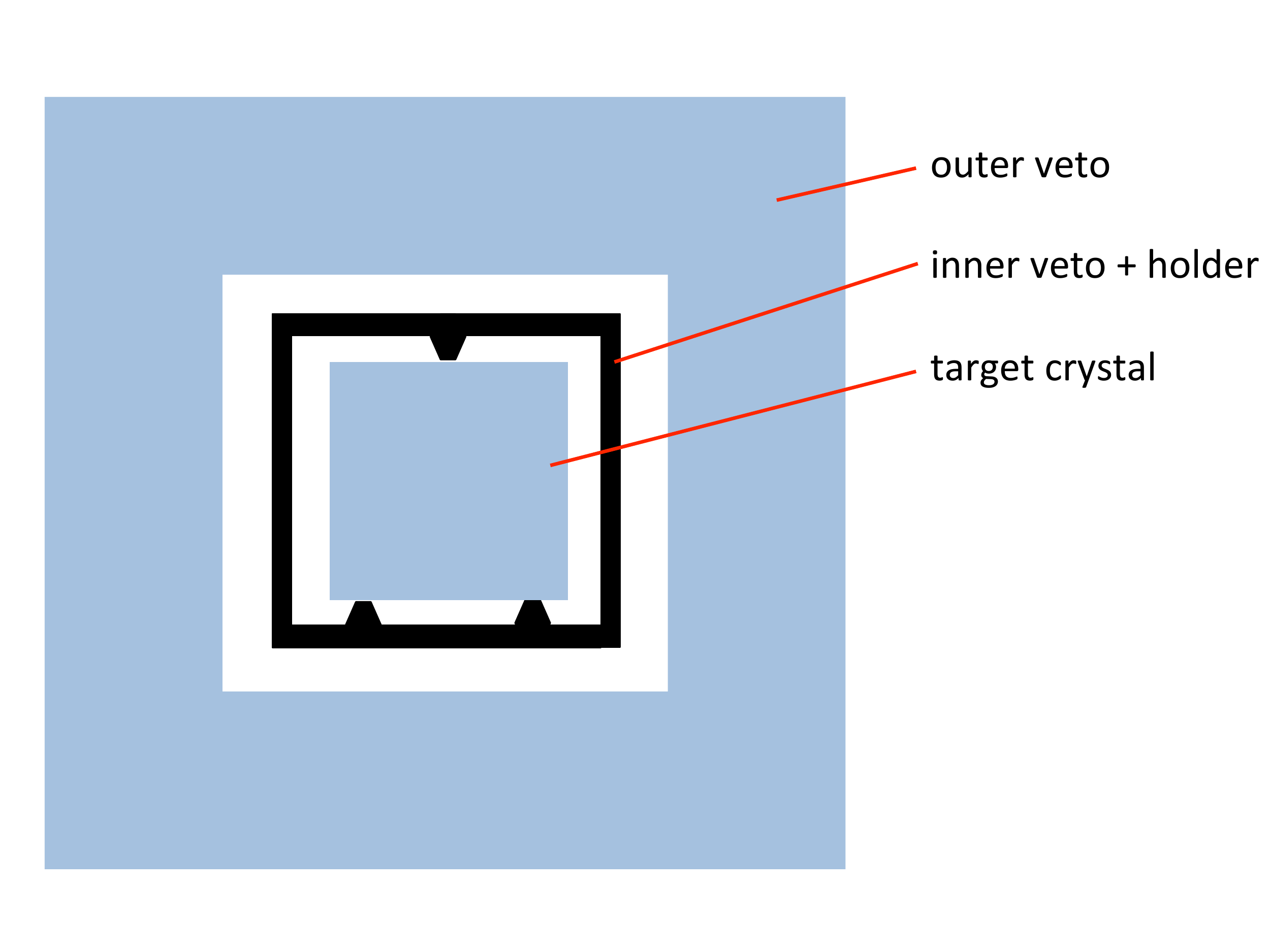}
\caption{Schematic view of the new detector which consists of 3 individual cryogenic calorimeters. The combination of, both, the outer veto against external gamma/neutron radiation,  and the inner veto against surface alpha and beta decays, significantly reduces the background level in the target crystal. In this way, a fiducial-volume cryogenic detector is realized.  The inner veto acts additionally as instrumented holder of the target crystal to reject possible stress-related relaxation events. }
\label{fig:concept}
\end{figure}

Here, a cryogenic detector is presented which realizes a fiducial volume by combining 3 individual calorimeters: 1) a target crystal (the fiducial volume) with an extremely low threshold of $\mathcal{O}$(${\lesssim}10$\,eV), 2) an inner veto as a 4$\pi$  veto against surface beta and alpha decays, and 3) an massive outer veto against external gamma/neutron radiation (see Fig. \ref{fig:concept}).  Additionally, the inner veto acts as an instrumented holder for the target crystal allowing to discriminate holder-related events (e.g. from stress relaxations). 

\subsection{Performance model for calorimeters}
\label{sec:performance}
In order to design the new detector,  a simple model was developed to  predict the practically achievable performance of  calorimeters of different geometry, material and mass\,\cite{rothe_master}. The model is based on experimental results of cryogenic CRESST-type detectors. The main results are derived here, insofar as they drive design-choices for the fiducial-volume cryogenic detector.


The fundamental equation describing a calorimeter is that, for a system in internal thermal equilibrium, the temperature rise
\begin{equation} \label{jump}
\Delta T= \frac{\Delta E}{C}
\end{equation}
where ${\Delta E}$ is an energy deposit and $C$ is the heat capacity
of the object. Reducing $C$ yields a large increase in temperature and thus a
high sensitivity to small energies.


Present cryogenic detectors  of $\sim 300$\,g achieve energy thresholds of $\sim300$\,eV\cite{Angloher:2015ewa}. In this work we investigate the performance and potential of gram-scale devices. 

The fundamental energy resolution $\sigma_E$ of cryogenic
calorimeters is given by  irreducible  thermal fluctuations between
the absorber and the thermal bath \cite{moseley_xray}: 
\begin{equation}
\sigma_E^2 \sim k_BT^2C 
\end{equation} 
with the absorber's temperature $T$, heat capacity $C$ and the
Boltzmann constant $k_B$. This corresponds to theoretical energy
resolutions of $\mathcal{O}$(1\,eV)  at ${\sim10}$\,mK for massive calorimeters with
masses of $\sim100$\,g  \cite{Pyle:2015pya}. Phonon processes in
cryogenic calorimeters  with thin-film transition-edge-sensors
(TES) as considered in this work are well described by a dedicated
thermal model \cite{Probst:1995fk}.

Equ. ~\ref{jump} is valid for a thermometer measuring the temperature of an absorber. 
In practice, the thermometers of cryogenic detectors can only measure their own temperature. 
Equ.~\ref{jump} thus changes to
\begin{equation}
\Delta T = E_{abs}/C_{film}
\end{equation} 
where $E_{abs}$ denotes the energy absorbed in the thermometer and $C_{film}$ the  heat 
capacity of the thermometer film. In cryogenic calorimeters at very low temperatures ($\sim 10$\,mK), the energy deposition in the thermometer film 
happens via the absorption of non-thermal phonons, which propagate ballistically and interact directly with the 
metallic film electrons. Thus they are not affected by the weak thermal coupling between thermometer phonon
and electron systems at such temperatures. 
To achieve sufficiently low heat capacities, temperatures as low as 10mK are required for these devices. The strong electron-phonon decoupling in the thermometer film at these temperatures requires a dedicated thermal link to the heat bath. This strongly suppresses the thermal signal, which makes the non-thermal phonon component our dominant information carrier. 

The thermometer's temperature rise can therefore be written as the ratio of the time-constants of the two competing
processes that reduce the non-thermal phonon population: 1) the absorption in the thermometer with a 
time-constant $\tau_{film}$, and 2) the thermalization of non-thermal phonons at the crystal surfaces with a 
time-constant $\tau_{c}$
\begin{equation}\label{timeconst}
\Delta T = \frac{\tau_{c}}{\tau_{film}}\cdot\frac{\Delta E}{C_{film}}.
\end{equation} 
It should be noted that this is only valid in the limit $\tau_{c}\ll\tau_{film}$, which is 
equivalent to the statement that collection by the thermometer film does not influence the non-thermal 
phonon lifetime~\cite{Probst:1995fk}. All devices considered here operate in this regime.
Under these conditions, the temperature signal is not influenced by the presence of the thermometer, and 
thermometer optimization can be considered separately from a change in absorber parameters.


For the absorber scaling law, we keep only the quantities that depend on absorber properties.
The energy threshold of the device is inversely proportional to the temperature rise, so we can write 
\begin{equation}
E_{th} \propto \frac{\tau_{film}}{\tau_{c}}.
\end{equation} 
This is the basis for our scaling law which only considers varying absorber material, geometry
and mass. Under these changes, $\tau_{c}$ scales with the average time between surface scatterings of 
the non-thermal phonons, which can be written
\begin{equation}\label{tauc}
\tau_{c}\propto\frac{l}{\langle v_g\rangle}
\end{equation} 
in terms of the mean phonon free path in the crystal $l$ and the mean phonon group velocity 
$\langle v_g\rangle$. For a fixed thermometer surface area, $\tau_{film}$ scales with the crystal volume and
the mode-averaged absorption rate, like
\begin{equation}\label{taufilm}
\tau_{film}\propto\frac{V}{\langle v_\perp\alpha\rangle}.
\end{equation} 
$v_\perp\alpha$ is the volume spanned by the phonon modes that cross the thermometer surface per unit time 
and thermometer area, times the transmission probability into the thermometer. The different dimensionality,
(i.e. $l$ vs. $V$), in the scaling laws, arises from the fact that the crystal surface area scales up with the
system dimensions, whereas the thermometer area does not.

In total, the scaling law is:
\begin{equation}\label{equ:scaling_law}
E_{th} \propto\frac{V }{l}\cdot\frac{\langle v_g\rangle}{\langle v_\perp\alpha\rangle}
\end{equation} 
The first part is purely geometric, while the second represents material parameters. The threshold  of CaWO$_4$ detectors is expected to be 1.72 higher than Al$_2$O$_3$ of same geometry \cite{rothe_master}, while Si (1.42) and Ge (1.15) fall between these two.
The scaling of two detector geometries as a function of mass are considered here. 
1)~For cubes of side length $d$, $V\propto d^3$ and $l\propto d$, so that $E_{th}\propto d^2$ which yields 
finally $E_{th}\propto m^{2/3}$.
2)~For plates of area $d^2$ and fixed thickness $h$, $V\propto d^2$. In the relevant range,
$10\lesssim d/h \lesssim 100$, $l(d)$ rises slowly from $\sim 2h$ to $\sim 5h$ (from MC simulation). Roughly,
we can take $l\approx \mathrm{const}$, which also gives $E_{th}\propto d^2$, but a different mass-scaling 
$E_{th}\propto m$. 

\begin{figure}
\centering
\includegraphics[width=0.45\textwidth]{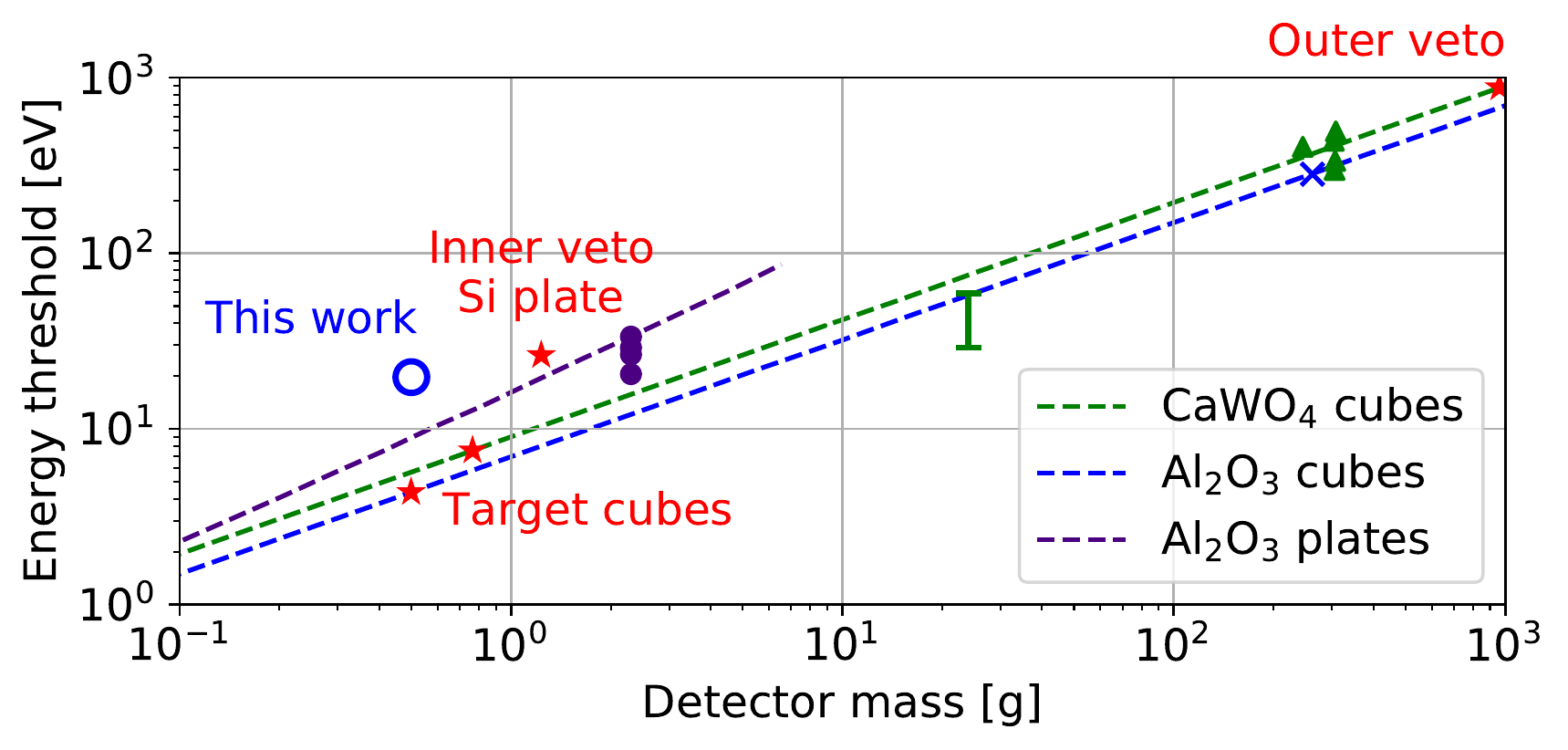}
\caption{Overview of the performance of different calorimeters: the nuclear-recoil energy threshold is plotted vs. the detector mass. The bands (1\,$\sigma$) show predictions of the performance model for CaWO$_4$ (green) and Al$_2$O$_3$ (blue) calorimeters (see main text). The model is fitted to data of existing CRESST-type detectors (green and blue) \cite{rothe_master,Angloher:2014myn,Angloher:2015ewa,Meier2000350,Angloher200243,Strauss2016}. Red stars indicate the predicted performance of the detector components studied in this work.  The prediction bands for the new Si calorimeters are not shown for clarity.}
\label{fig:predictions}
\end{figure}

{\color{MPPgreen} }
With values for $l$ found by Monte Carlo methods for each occurring detector geometry, the model can be used to describe the thresholds of various CRESST-type detectors. Since the model can only predict a scaling under change of absorber properties, the absolute normalization (depending e.g. on the noise level of the setup) has to be taken from the respective experiment. In the following, the noise level of the "benchmark" CRESST setup at LNGS is considered. 
 In Fig. \ref{fig:predictions}, the model predictions for plate and cube detectors are shown as a function of detector mass, fitted to the thresholds achieved in CRESST-II CaWO$_4$ detectors (green triangles) with a mass of $\sim300$\,g\,\cite{Angloher:2014myn,Angloher:2015ewa} and a sapphire cube of 262\,g used in CRESST-I (blue cross)\,\cite{Meier2000350,Angloher200243}.  
The model successfully predicts the energy threshold of CRESST-II light detectors studied in \cite{rothe_master} (purple dots), which are sapphire discs with a mass of 2.2\,g (diameter: 40\,mm, thickness: 0.45\,mm) and also the thresholds  of ${\sim}24$\,g CRESST-III  detectors as expected from a prototype measurement (green error bar) \cite{Strauss2016}. 
The capability to extrapolate calorimeter thresholds for different detector geometries and materials over orders of magnitude in mass can be applied to the component design for the fiducial-volume cryogenic detector.
Red stars indicate the calculated performance of the calorimeters studied here.  

\subsection{Design of the target calorimeter array}\label{sec:target_cal}
\begin{figure}
\centering
\includegraphics[width=0.45\textwidth]{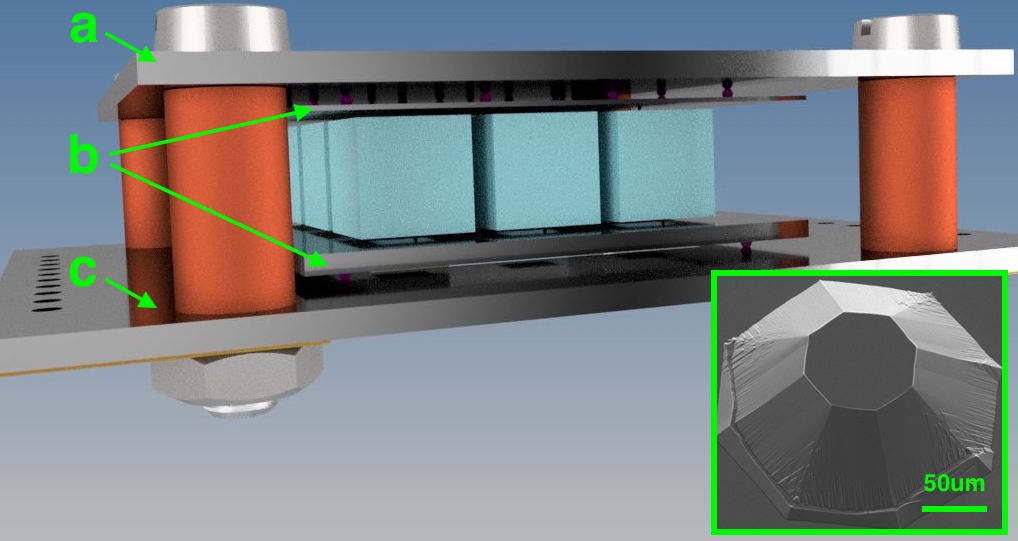}
\caption{Technical drawing of the calorimeter array. 3x3 cubic calorimeters (e.g. CaWO$_4$, Al$_2$O$_3$) are installed between two dedicated Si  wafers (b). The contact area to the cubic crystals is realized by pyramides (height $200\,\mu$m) which are produced by wet chemical etching. The inset shows a microscopic picture of a prototype pyramidrical structure.  The outer Si wafers (a,c) act as holding structure and host the electrical wiring.  }
\label{fig:array_picture}
\end{figure}

For the research program proposed in this work, the target calorimeter has to fulfill the following requirements:
\begin{itemize}
\item A nuclear-recoil energy threshold $E_{th}$ of $\mathcal{O}(10$\,eV). 
\item  Rates of ${10^2-10^3}$/[kg day] are expected from CNNS at the sites studied here, as will be shown in chapter \ref{sec:CNNS}. Corresponding to this rate, a total target mass of ${\sim}10$\,g is needed for the detection of CNNS. 
\item Lowest thresholds require a sufficiently low event rate in the calorimeter. To limit the pile-up contribution to a level of $\mathcal{O}$($10^{-2}$), a maximum event rate of $\mathcal{O}$(0.1\,Hz) per detector is allowed given the typical (thermal) pulse decay times of ${\sim}100\,$ms \cite{Probst:1995fk}. 
\item The CNNS cross-section is proportional to the target's neutron number $N$ squared, which highly favors heavy elements. On the contrary, the use of light nuclei facilitates a characterization of neutron backgrounds. 
\end{itemize}
Considering these design features, cubic target crystals with a edge length of 5\,mm equipped with a tungsten thin-film TES are ideal. A multi-target approach with a variety of elements is chosen which has great advantages for the separation of signal and background through characteristic interaction strength.   Cubes of CaWO$_4$ (0.76\,g), Al$_2$O$_3$ (0.49\,g), Ge and Si crystals, which are well-known for their excellent cryogenic detector properties \cite{Angloher:2015ewa,Angloher200243},  are suitable candidates.  The performance model (see chapter \ref{sec:performance}) predicts energy thresholds of $E_{th}{\approx}4.0$\,eV for Al$_2$O$_3$ and $E_{th}{\approx}7.0$\,eV for CaWO$_4$ (see red stars in Fig. \ref{fig:predictions}). To obtain the desired total target mass, a 3x3 detector array is foreseen (see Fig. \ref{fig:array_picture}).  This corresponds to a total target mass of 6.84\,g for the CaWO$_4$  and 4.41\,g for the Al$_2$O$_3$ array, respectively.  

\begin{figure}
\centering
\includegraphics[width=0.45\textwidth]{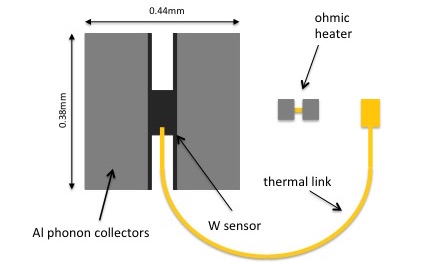}
\caption{Schematic view of the sensor design for the target calorimeter. A thin W film (thickness 200\,nm) weakly coupled to the heat sink via a  (0.01x7.0\,mm$^2$ thickness 20\,nm) Au stripe. An Al phonon collector is attached to the W film to increase the collection area of the sensor. The readout current on the TES,  the signals on the separate ohmic heater, and the thermal contact are connected via Al and Au wire bonds, respectively.}
\label{fig:TES}
\end{figure}

For the temperature sensor, a  TES is chosen similar to that which is used for the CRESST detectors \cite{Strauss2016}. The TES consists of a thin W film (thickness 200\,nm) with an area of 0.0061\,mm$^2$ and an Al phonon collector with an area of 0.15\,mm$^2$ attached to it (see Fig. \ref{fig:TES}). The latter increases the collection area for phonons without the penalty of increasing the heat capacity of the sensor \cite{Angloher2016} yielding an increased pulse height. The TES is weakly coupled to the heat sink via a thin Au stripe (0.01x7.0\,mm$^2$, thickness: 20\,nm)  providing a thermal conductance of ${\sim}10\,$pW/K (at a temperature of 10\,mK). Al and Au wire bonds with a diameter of 25\,$\mu$m are used to provide the electrical contacts for the TES (bonded on the phonon collectors) as well as the ohmic heater (separate bond pads), and the thermal link to the heat sink, respectively. Typically, bias currents between 100\,nA and 5\,$\mu$A   are applied to the sensor. The resistance change of the TES can be measured with a  SQUID system similar to the one in the CRESST dark matter experiment \cite{Angloher:2012vn}. 

\subsection{Results from a prototype calorimeter}\label{sec:prototype}
In the framework of this project, a prototype Al$_2$O$_3$ calorimeter of 0.5\,g has been produced and equipped with a TES according to the design goals described in the previous section. The detector was installed in a copper holder and mounted in a  detector test facility at the Max-Planck-Institut for physics in Munich. It consists of a dilution refrigerator in a surface building without dedicated shielding against ambient radioactivity. Further, no shielding against backgrounds from surfaces in the direct vicinity of  the calorimeters is used. A $^{55}$Fe X-ray source is placed close to the detector for a calibration of the low-energy region. 

In an accompanying paper \cite{cnns_letter}, we present details of a 5.1\,h calibration measurement performed with the 0.5\,g Al$_2$O$_3$ detector which achieved an energy threshold of $E_{th}=(19.7\pm 0.9)$\,eV. This is  independent of the type of particle interaction  since it is a calorimetric device. This is the lowest nuclear-recoil energy threshold  reported for massive calorimeters, beyond the fundamental nuclear-recoil reach of ionisation-based detectors \cite{Agnese:2016cpb}. 

The detector operates in the calorimetric mode (see section \ref{sec:performance}), confirmed by the pulse shape. The thermalisation times in the crystal and thermometer film are found to be $\tau_{c}=0.34$\,ms and $\tau_{film}=2.2$\,ms, respectively. This ratio fulfills the condition $\tau_{c}\ll\tau_{film}$ but leaves room for improvements (see below).

 The  measured threshold is higher (by a factor of $\sim$5) compared to what is predicted by the performance model for calorimeters (section \ref{sec:performance}). Part of the discrepancy  may be explained due to worse noise level in the MPI setup (by factor 1.5-3 \cite{Strauss2016}) compared to the low-noise benchmark setup  used for the calculation of the predictions (Equ. \ref{equ:scaling_law}). The considered detector, being the first prototype of a gram-scale calorimeter, is expected to  improve by further developments and adjustments of the TES sensor. The ratio of $\tau_{c}/ \tau_{film}$  can be further decreased by reducing the thermometer area and accordingly weakening the thermal link. A corresponding reduction of the Al  phonon collectors may improve the transport efficiency of quasi particles \cite{Angloher2016}.  Furthermore, a moderate reduction of the W-film thickness will reduce the heat capacity of the thermometer without compromising the phonon absorption. 

In the calibration measurement a flat background spectrum of $\sim1.2\cdot10^{5}$\,counts/[kg keV day] (7-10\,keV) is observed above the calibration peaks \cite{cnns_letter}. This  is expected due to the absence of any shielding and can be considered as an absolute upper limit for the total rate in surface experiments (here it corresponds to $\sim0.3$\,Hz). It is comparable to typical total acount rates observed in $\mathcal{O}$(1\,kg) cryogenic detectors operated in underground laboratories \cite{Angloher:2014myn}. The  result clearly demonstrates that gram-scale detectors can be operated in a high-background environment -- in particular at surface level -- while allowing low energy thresholds and stable operating conditions. 

The performance of the prototype fulfills the  requirements  listed in the previous section in terms  energy threshold and operability at surface-level. To demonstrate the required background level -- for the near future -- measurements with further developed CaWO$_4$ and  Al$_2$O$_3$  calorimeters at low-background experimental sites (e.g. a shallow laboratory) are planned. In particular, the target calorimeter(s) will be embedded in the inner and outer cryogenic shieldings which are discussed in detail in the following sections.

\subsection{Low-threshold inner veto and detector holder}\label{sec:innerVeto_holder}
Background from the surfaces of the target crystals and surrounding surfaces is a big challenge for rare-event searches, and can limit the sensitivity at low energies. The inner veto provides an active discrimination against beta and alpha decays occurring on surfaces. Typical Q-values of such decays are between $\sim10$\,keV and $10\,$MeV typically shared between 2,3 or more product particles leaving the interaction point in different directions. In a configuration where the target is surrounded by a $4\pi$ active veto, the total energy of the reaction is detected (apart from the energy transferred to neutrinos in beta decays). In this way, a high fraction of such backgrounds can be rejected by coincident events in the veto.  The rejection of surface background is crucial in particular when approaching ultra-low energy thresholds, as can be seen in experimental data (see e.g.\cite{Angloher200243}). Fig.  \ref{fig:innerveto_schematic} shows a section view  of the inner part of the detector. In the following, the functionality of the relevant detector components is briefly discussed:
\begin{figure}
\centering
\includegraphics[width=0.45\textwidth]{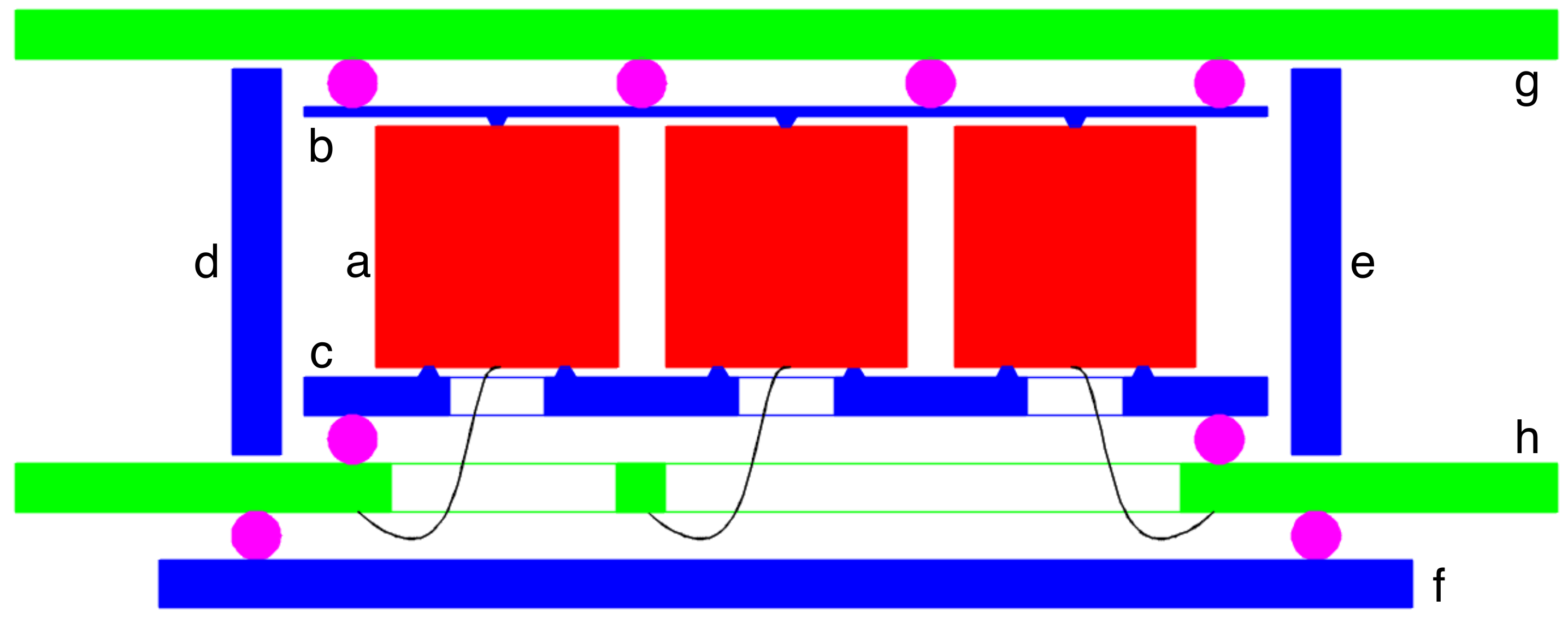}
\caption{Schematic side view of the detector array.  Red: target calorimeter cubes (a) of (5x5x5)\,mm$^3$ with a TES each. Blue:  Si wafers instrumented with TESs providing a 4$\pi$ surface veto (b,c,d,e,f). Two slabs have pyramides (b,c) with a height of $200\,\mu$m which are produced by wet chemical etching. These structure holds the target crystals.  Slab b is flexible due to a thickness of only 200$\,\mu$m.      Purple: Sapphire balls with a diameter of 1\,mm.  Green: Two outer Si slabs (g,h) of 2\,mm thickness press together the inner part. Slab b thereby acts as a spring.  Slab h  hosts the electrcal wiring which is connected to the  TES  with  wire bonds (black).  }
\label{fig:innerveto_schematic}
\end{figure}
\begin{itemize}
\item Target (red): The detector consists of 9 target calorimeters (a in Fig. \ref{fig:innerveto_schematic}) arranged in a 3x3 detector array.  Each crystal is equipped with a TES (see chapter \ref{sec:target_cal}). 
\item Active components (blue): To realize a 4$\pi$ veto against surface backgrounds,  Si wafers read-out by a TES each are used  (b-f). Two of these (b and c) are in contact with the target crystals via pyramidal Si structures on the wafers. The upper one (b) is thin enough  ($200\,\mu$m) to be flexible - the wafer acts as a spring. Pressed to the target crystals, the thin wafer realizes a spring-loaded holding structure which can compensate for thermal contraction of the various components of the detector. Possible events induced by the detector holder (e.g. by thermal-stress relaxation) can be rejected since they induce also  phonon signals in the TESs of b and c.  
\item Passive components (green): Two Si slabs (g and h) are used as support structures for the calorimeter array. They are attached to each other by 4 posts (shown in Fig. \ref{fig:array_picture}) providing the necessary pressure to hold the target crystals. The lower wafer (h) is equipped with  Al (Au) wiring for the electrical (thermal) connection of the target calorimeters and the inner veto devices.  
\end{itemize}

The inset in Fig. \ref{fig:array_picture} shows a prototype Si wafer with a pyramid structure produced  at the Halbleiterlabor of the Max-Planck-Society. The structure is defined by photo-lithography techniques and the pyramid structures are then realized by wet chemical etching. 

The rejection power against surface related background was estimated with a
dedicated Monte Carlo (MC) study performed with the Geant4 code in version
10.2p1 \cite{g4a,g4b}. We follow the recommendation of the \textit{Geant4 Low
Energy Electromagnetic Physics Working Group} \cite{g4c} and implement the low
energy behaviour of electromagnetic interactions via the Geant4 class
\textit{G4EmStandardPhysics\_option4}, a selection of most accurate models.
Furthermore, we enabled the atomic de-excitation via emission of fluorescence
photons and Auger electrons. With one exception, we applied a production cut
 of $250\,\mathrm{eV}$ throughout our geometry, i.e.\ for energies above this
 cut new secondary particles can be created in the simulation, whereas lower
 energies are directly deposited. The exception are fluorescence photons and
 Auger electrons which are produced in any case.

\begin{figure}
\centering
\includegraphics[width=0.5\textwidth]{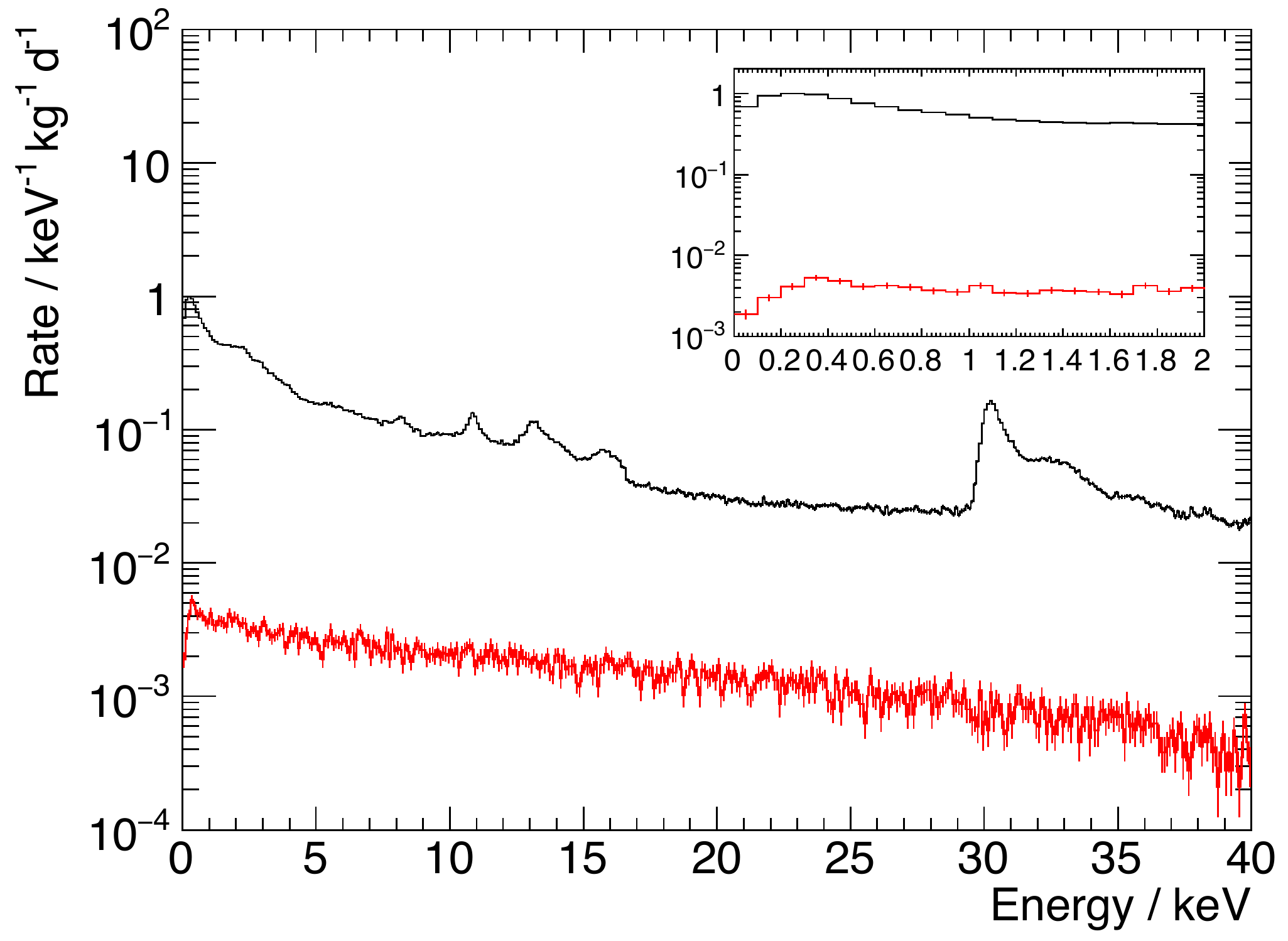}
\caption{MC simulation of the expected background from a contamination of 
the inner surface of the inner veto with a surface $\beta$-emitter
($\mathrm{^{210}Pb}$). The histograms shows the energy deposits in the target
for two cases: a passive inner veto (\textit{black}) and an active inner veto
(\textit{red}) with a threshold of $30\,\mathrm{eV}$. 
The \textit{inset} zoom to the first $2\,\mathrm{keV}$. 
Clearly a background 
reduction of $\mathcal{O}(10^2)$ is feasible at low energies}
\label{fig:innerveto_simulation}
\end{figure}

Exemplary for an surface contamination we simulated the $\mathrm{\beta}$-decay
of $\mathrm{^{210}Pb}$ by placing the lead ions at rest on the inner surface of
the inner veto, facing one target calorimeter made of $\mathrm{Al_2O_3}$. The
source activity is assumed to be
$\mathcal{O}(1\,\mathrm{kg^{-1}keV^{-1}d^{-1}})$, the maximal external
$\beta$-activity observed with TUM40, a module with especial low background
operated in CRESST-II phase 2 \cite{strauss:2014part2}. The black histogram in
Fig.~\ref{fig:innerveto_simulation} shows the background spectrum seen by the
target with inactive inner veto, the red histogram shows the spectrum of the
remaining background in case of an active veto with a threshold of
$30\,\mathrm{eV}$.
Clearly a reduction of more than two orders of magnitude is feasible. 
A more 
detailed MC study of the complete detector array is underway and intended for future publication.

We note that the step at $\sim 100\,\mathrm{eV}$
(Fig.~\ref{fig:innerveto_simulation}, \textit{black histogram}) is no artefact
of the used production cut. Instead, it is caused by Coster-Kronig
transitions as part of the atomic relaxation subsequent to the decay of the $46.539\,\mathrm{keV}$-level of
$\mathrm{^{210}Bi}$ to which $\mathrm{^{210}Pb}$ decays in 84\% of
the cases \cite{firestone}.


\subsection{Outer-veto detector}\label{sec:outer_veto}
\begin{figure}
\centering
\includegraphics[width=0.48\textwidth]{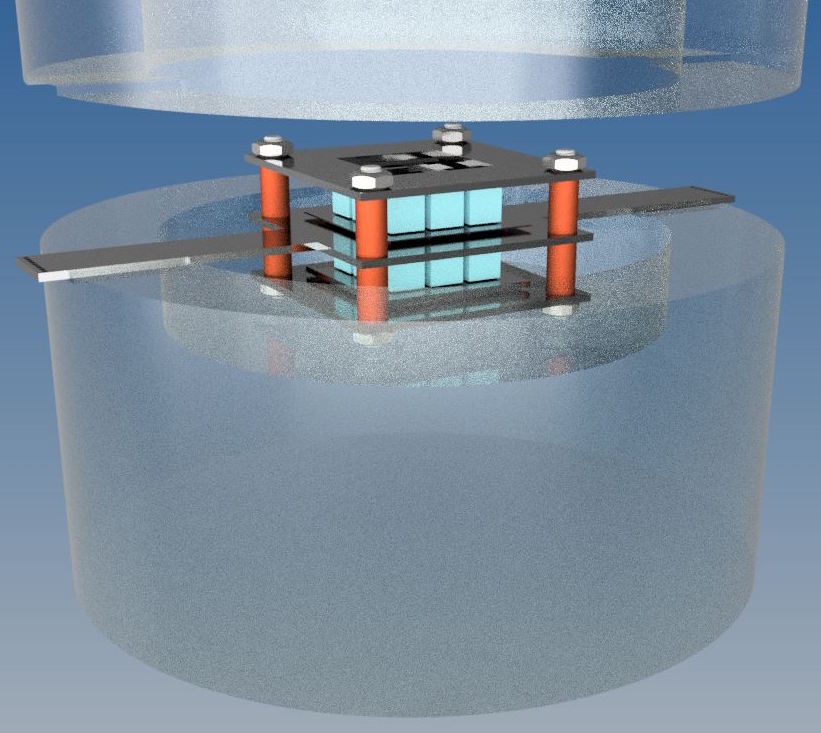}
\caption{Technical drawing of the fiducial volume detector. Two calorimeter arrays are installed inside the  CaWO$_4$ outer veto with diameter of 10\,cm. The veto is made of two parts with a height of 5\,cm each which are equipped with TESs and operated as cryogenic detectors.  }
\label{fig:outerveto}
\end{figure}

Given the smallness of the calorimeter array and the inner veto system, these components can be embedded in a large cryogenic outer veto. We consider cylindrical crystals with a diameter and height of $\mathcal{O}$(10\,cm) which are segmented into two (or more) parts with a central cavity to host the inner detector parts (see Fig. \ref{fig:outerveto}). Each crystal of the outer veto is instrumented with a TES. It is foreseen to use materials that are known for their excellent phonon properties, such as e.g. Ge and CaWO$_4$, and that have been demonstrated as cryogenic detectors with masses of $\mathcal{O}$(100\,g-1\,kg). Thresholds between 300\,eV and 1\,keV are reached with such devices, in agreement with the prediction of the performance model for calorimeters in section  \ref{sec:performance} (Fig. \ref{fig:predictions}).
CaWO$_4$ is the preferred material: it has the heavy element W which provides a high cross-section for gamma radiation and the relatively light element O for an efficient moderation of neutrons. The simulations below are therefore performed using CaWO$_4$. However, when scaling up the number of detectors (see section \ref{sec:scalability}) larger diameters of CaWO$_4$ crystals are necessary which currently are not available. In this case, Ge crystals are a promising alternative, since those are readily produced in large diameters (up to 300\,mm), with high radiopurity.

It is worth mentioning that the timing information of pulses in the cryogenic detectors is crucial for the efficiency of a coincidence veto. The precision with which the onset of the pulses can be determined defines the dead time in the target calorimeter. We know from neutron scattering experiments that the pulse onset of comparable cryogenic calorimeters can be determined with a uncertainty of $\pm5\,\mu$s \cite{Strauss2014}. Even an excessive rate of 100\,Hz in the veto detector would introduce only a negligible deadtime of $\lesssim$0.1\%.

\begin{figure}
\centering
\includegraphics[width=0.48\textwidth]{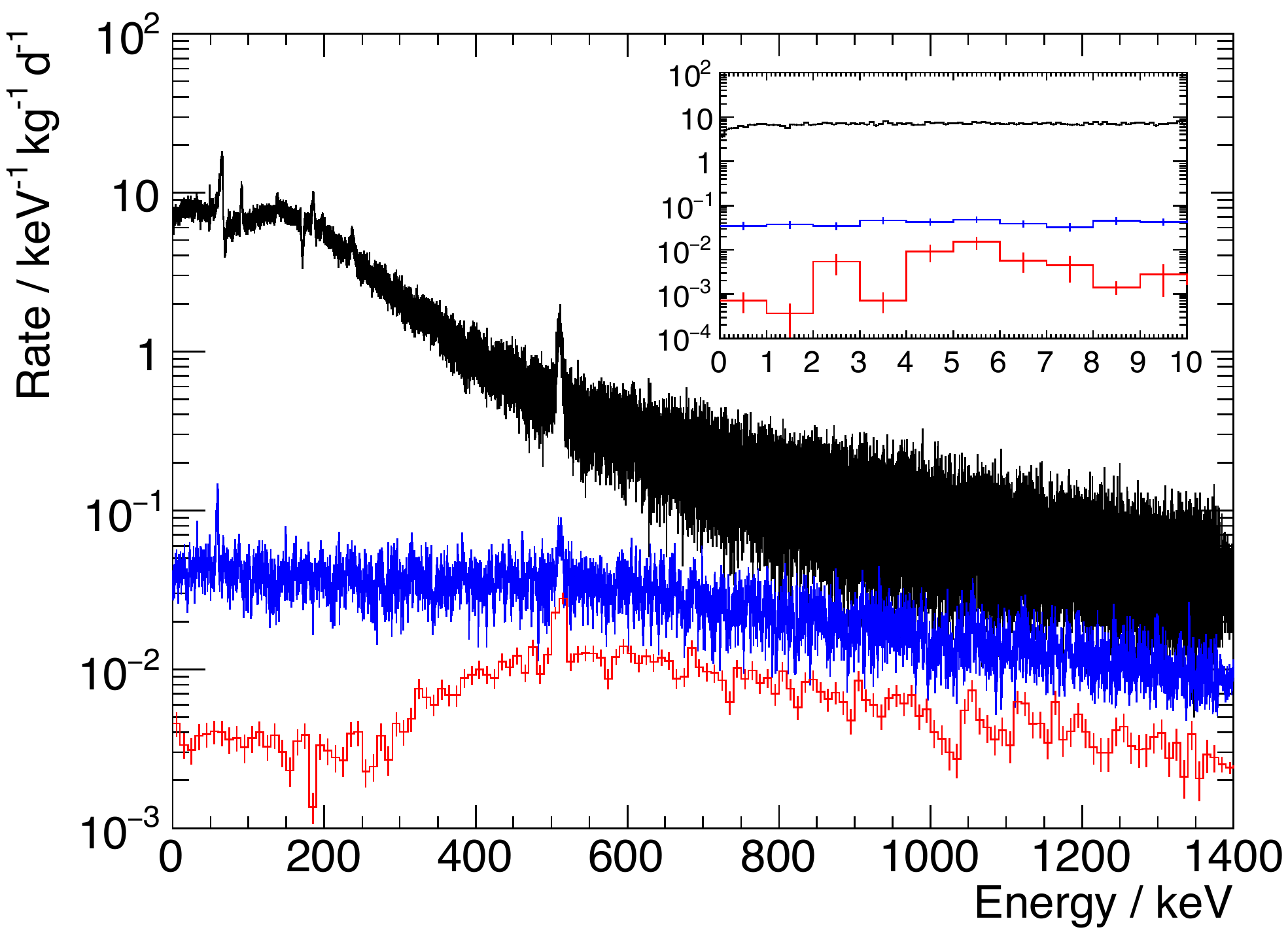}
\caption{MC simulation of the expected energy deposit in case of a $\gamma$-background similar to the remaining one in the Dortmund Low Background facility \cite{Gastrich:2015owx}. The histograms show the energy deposits in the target for three cases: without any veto (\textit{black}), in case of a passive outer veto (\textit{blue}), and in case of an active outer veto (\textit{red}) with a threshold of $1\,\mathrm{keV}$. The \textit{inset} zoom to the first $10\,\mathrm{keV}$. Clearly a background reduction of $\mathcal{O}(10^3)$ at lowest energies is reasonable.}
\label{fig:outerveto_simulation}
\end{figure}

Also the rejection power of the outer veto was estimated with a MC study. Here a $\mathrm{CaWO_4}$ target was placed inside the nested shields of inner and outer veto. As typical background we investigate gamma rays following the remaining spectrum at the Dortmund Low Background facility \cite{Gastrich:2015owx}, a low-background site at the surface which will be discussed in section~\ref{sec:CNNS}. FIG.~\ref{fig:outerveto_simulation} shows as black histogram the background spectrum observed by an unshielded target, in blue the remaining background in case of a passive outer veto, and in red the remaining background in case of an active outer veto with a threshold of $1\,\mathrm{keV}$. Even with only a passive veto  a background supression of more than 3 orders of magnitude at low energies is feasible. Activating the outer veto increase the supression to more than 4 orders of magnitude. Importantly, the expected gamma-induced electron-recoil spectrum remains flat down to energy threshold (see inset of Fig. \ref{fig:outerveto_simulation}). 

For a first estimate of muon-induced neutron backgrounds, a basic MC simulation was performed. Using an active  CaWO$_4$ outer veto, the neutron background is reduced by a factor of $\sim10$, independent of the recoil energy (studied in the energy range from 10\,eV to 300\,keV). By a clever combination of passive shielding elements like borated polyethylene, and active shielding elements like instrumented plastic or liquid scintillators, and LiF crystals,  neutron background levels can be further reduced. This concerns shielding systems placed outside the cryogenic setup surrounding the cryostat at all sides.  In addition we provide two technologies to further reduce and reject this potentially harmful background: 1) the outer cryogenic veto system described above and 2) the active background discrimination by the multi-target approach. The latter might be a powerful tool to reduce ultimate backgrounds, particularly neutrons. This is described in more detail in section \ref{sec:CNNS}. Nevertheless, we conclude that a dedicated MC simulation using measured muon spectra in combination with a calorimeter measurement at the experimental site are necessary. This is beyond the scope of this work and will be subject of a future publication.

\subsection{Production and scalability}\label{sec:scalability}

A disadvantage of cryogenic detectors when compared to e.g., scintillation detectors has always been the difficulty to scale up the experiments  in size. The new detector concept presented here overcomes most of these problems. In principle, the detector has been designed such, that the number of production steps of the individual  detector components are independent of the number of target calorimeters involved.

 The target calorimeters are produced from wafers with a thickness of 5\,mm and variable diameters (CaWO$_4$ up to 60\,mm, Al$_2$O$_3$ up to 200\,mm, and Si up to 300\,mm). With well-established techniques of the semiconductor industry, as e.g. photolitography, thin-film evaportation, etching or sputtering,  the TES sensors are being simultaneously equipped on each target calorimeter, and  the wafer is cut only afterwards into the individual (5x5x5)\,mm$^3$ crystals. The same up-scaling is possible for the inner veto (section \ref{sec:innerVeto_holder}) which acts as a  detector holder. It is entirely produced by the above-mentioned methods.  The cutting of the wafers is done by means of a laser or other automated methods.  The cabling for a large amount of  TES sensors are implemented by photolitography  in combination with sputtering on the inner veto wafers as done for the 3x3 array. Further, it has been shown (e.g. in \cite{2012RScI...83g3113D}) that large amounts of SQUIDs can be realized by SQUID multiplexing. 
 
\begin{figure}
\centering
\includegraphics[width=0.45\textwidth]{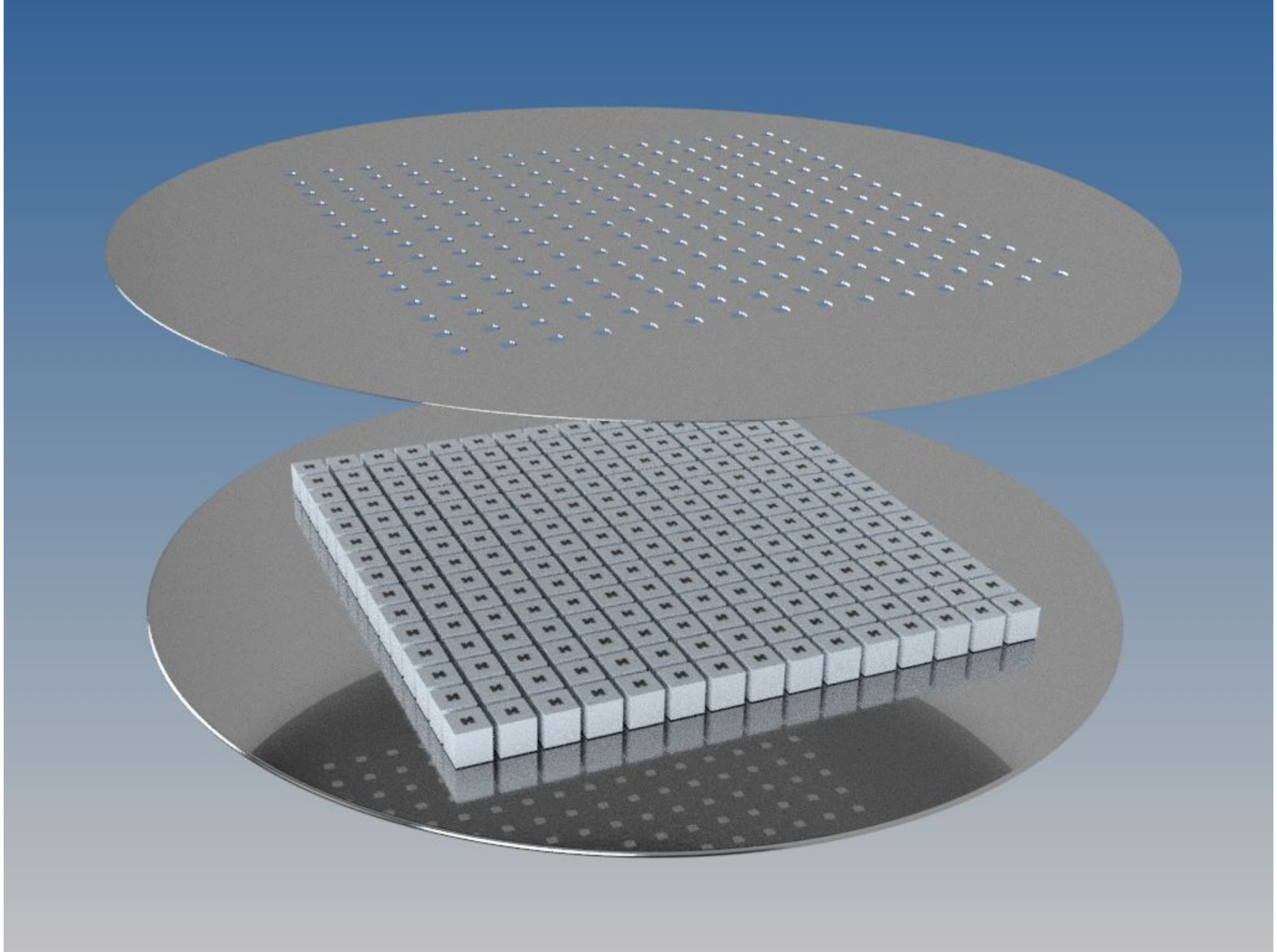}
\caption{Technical drawing of  a up-scaled calorimeter array using state-of-the-art wafer sizes of 150\,mm diameter (e.g. Al$_2$O$_3$ as target and Si as holder). In one production step, a total target mass of $\sim110$\,g can be achieved using an array of 225 crystals.}
\label{fig:upscale}
\end{figure}

 For the first phase of the experiment, we focus on the production of 3x3 arrays with moderate requirements of size and channel numbers which is foreseen as sufficient for a discovery of CNNS (see below). In a second step, the technology mentioned above enables  experiments up to the kg-scale with energy thresholds of $\mathcal{O}$(10\,eV); an exposure allowing precision measurements of the CNNS cross-section and interesting BSM physics.   Fig.~\ref{fig:upscale} shows a technical drawing of a future calorimeter array of 225 crystals which correspond, using  Al$_2$O$_3$, to a total mass of $\sim110$\,g. 

\section{Detection of Coherent Neutrino Scattering}\label{sec:CNNS}

\subsection{Case 1: At a nuclear power reactor}

\subsubsection{Signal expectation}

Nuclear power reactors are among the most intense (anti-) neutrino sources on earth and therefore a highly interesting site for the detection of CNNS. 

\begin{figure}
\centering
\includegraphics[width=0.5\textwidth]{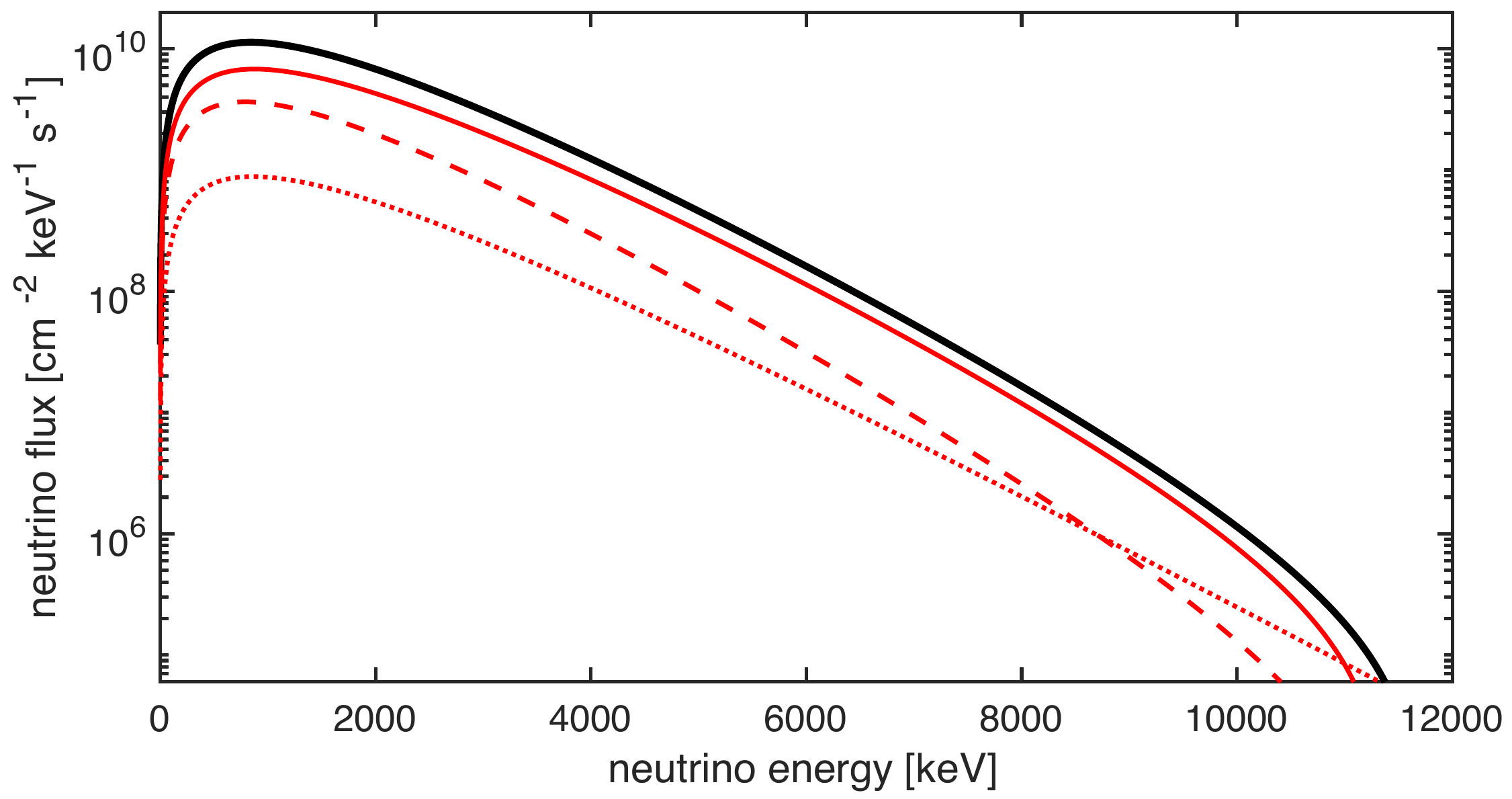}
\caption{Anti-neutrino flux from a benchmark pressurized-water nuclear reactor with a thermal power of 4\,GW at a distance of 15\,m.  A standard fuel composition is used for the calculation: 62\% of $^{235}$U, 30\% of $^{239}$Pu and 8\% of $^{238}$U. \cite{1989NuPhA.503..136T} }
\label{fig:flux_chooz}
\end{figure}

A benchmark reactor with a thermal power of 4\,GW, a typical value for a two-core reactor plant,  yields \mbox{$\sim 1.2\cdot 10^{20}$} fissions per second and an isotropic neutrino rate of $R_\nu\approx 7.5 \cdot 10^{20}$\,s$^{-1}$ \cite{Kopeikin:2004cn}. The neutrino flux $\Phi(E_\nu)$ can be calculated as
\begin{equation}
\Phi(E_\nu)=\frac{R_\nu}{4\pi d^2} \sum_{i} n_i \Phi_i(E_\nu) 
\end{equation}
 with the distance to the core $d$,  the fraction $n_i$ of the fuel component $i$ and the respective normalized neutrino-energy spectrum  $\Phi_i(E_\nu)$. Fig. \ref{fig:flux_chooz} shows the neutrino flux for a standard fuel composition (62\% of $^{235}$U, 30\% of $^{239}$Pu and 8\% of $^{238}$U \cite{1989NuPhA.503..136T}) from a 4\,GW reactor at a distance of $d=15$\,m from the core. 
The differential recoil spectrum in the detector can be written as 
\begin{equation}
\frac{dS}{dE_R}=N_t\int_{E_{min}}^{\infty}\frac{d\sigma(E_\nu,E_R)}{dE_R}\Phi(E_\nu)dE_\nu
\end{equation}
using Equ.\,\ref{equ:sigma}. $N_t$ is the number of target nuclei and $E_{min}=\sqrt{E_RM/2}$ the smallest neutrino energy leading to a recoil of a nucleus with the mass $M$.

\begin{figure*}
\centering
\includegraphics[width=0.9\textwidth]{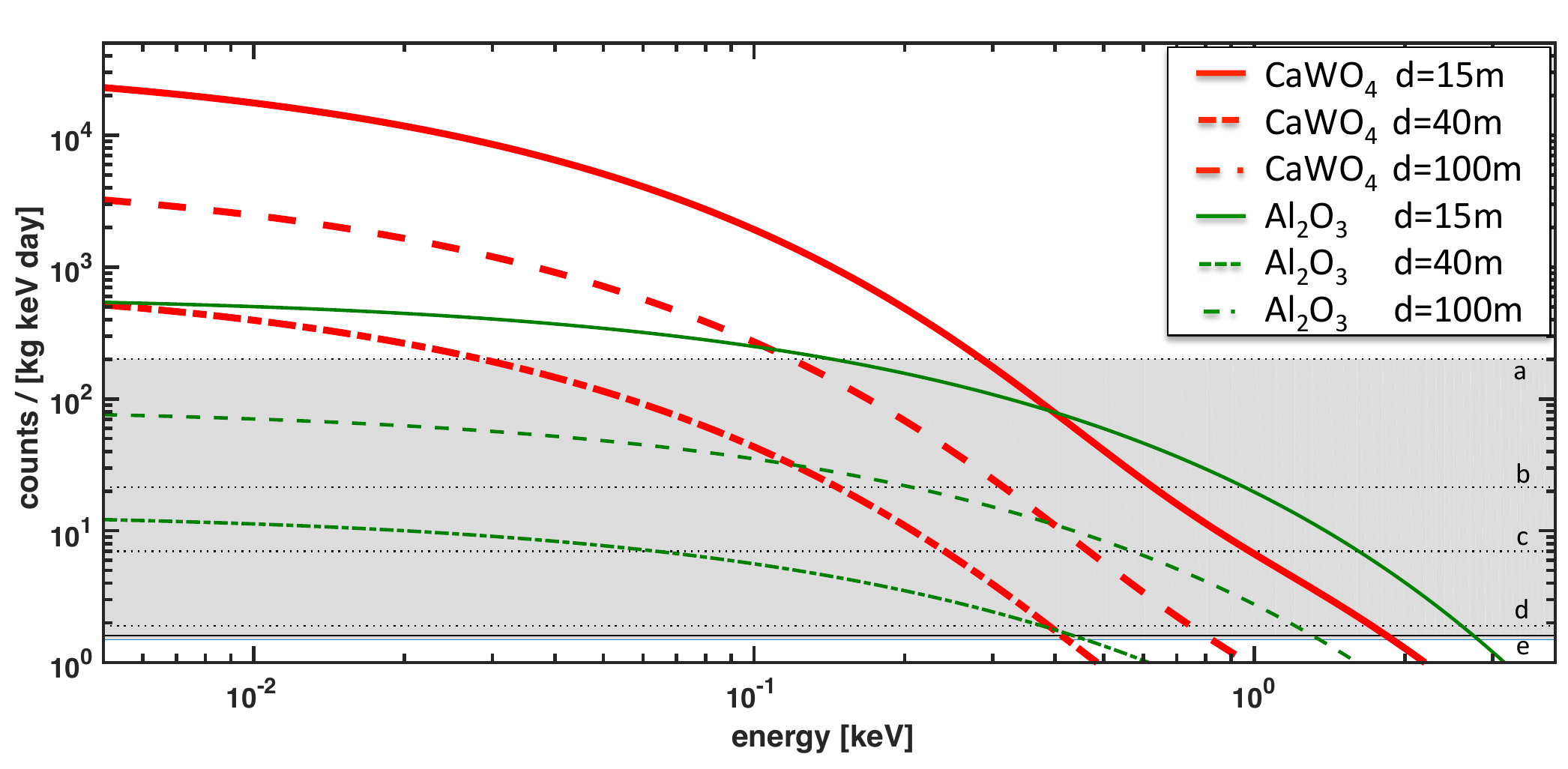}
\caption{Count rates  on CaWO$_4$ (red) and Al$_2$O$_3$ (green) expected from a benchmark nuclear power plant of 4\,GW for the 3 experimental sites considered. The black dotted lines indicate different background levels (extrapolation to lower energies) measured in different experimental sites. From top to bottom: a) the Stanford shallow underground facility \cite{Akerib:2010pv}, b) the low-background setup at the ARC in Seibersdorf \cite{Schwaiger2002375}, c) the Dortmund low-background facility \cite{Gastrich:2015owx} and d) the Heidelberg shallow laboratory \cite{Heusser:2015ifa}. The full black line (e) shows the expected (simulated) background level using the outer and inner veto of the fiducial-volume cryogenic detector.  The grey band indicates the uncertainty of the background level with a lower limit at the intrinsic background level of CaWO$_4$ crystals measured at LNGS \cite{strauss:2014part2}. Reactor-correlated backgrounds are considered as negligible at the considered distances from core. }
\label{fig:spectrum_chooz}
\end{figure*}

The differential recoil spectra of coherently scattered anti-neutrinos in CaWO$_4$ and Al$_2$O$_3$ detectors at different distances $d$ from the core of the benchmark reactor plant are shown in Fig. \ref{fig:spectrum_chooz}. Due to the $N^2$ dependency of the CNNS cross section (see Equ. \ref{equ:sigma}), the heavy element W  boosts significantly the rate on CaWO$_4$  below 100\,eV (full red line) to  $\sim4\cdot10^{4}$/[kg keV day]. 
The rate expected for Al$_2$O$_3$ (full green line), however, stays almost constant at a value of $\sim1\cdot 10^{3}$/[kg keV day] below $\sim300\,$eV. The rates for $d=40\,$m are about a factor of 7 lower (dashed lines).  The strong material dependence of the CNNS rate is  a powerful tool to discriminate the signal from irreducible backgrounds.  The signal rate is significantly different for CaWO$_4$ and  Al$_2$O$_3$, e.g. at 10\,eV the ratio is ${\sim9.3}$. In contrast, the  background counts from external gamma radiation is comparable (within a factor of $\sim2$). Further, similar neutron background spectra are expected since in both materials - for neutron induced scatters - dominantly O scatters are above energy threshold due to kinematics. 

\begin{table}
\caption{\label{tab:counts_chooz}Integrated CNNS count rate from a nuclear reactor with a total thermal power of 4\,GW at different distances $d$ between $E_{th}$ and 5\,keV. The rates are integrated up to 5\,keV.}
\begin{tabular}{lccccc}
\hline\noalign{\smallskip}
d [m]&$E_{th}$ [eV]&\multicolumn{2}{c}{counts/[kg day]}&\multicolumn{2}{c}{counts/[array day]}\\

&&CaWO$_4$&Al$_2$O$_3$&CaWO$_4$&Al$_2$O$_3$\\

\noalign{\smallskip}\hline\noalign{\smallskip}

15&5&790.3&112.8&5.44&0.51\\
&10&690.2&110.1&4.75&0.49\\
&20&547.2&105.4&3.77&0.47\\
40&5&111.1&15.9&0.77&0.07\\
&10&97.1&15.5&0.67&0.07\\
&20&77.0&14.8&0.53&0.07\\
100&5&17.8&2.5&0.12&0.01\\
 &10&15.5&2.5&0.11&0.01\\
 &20&12.3&2.4&0.08&0.01\\

\noalign{\smallskip}\hline
\end{tabular}
\end{table}
The integrated count rates for different energy thresholds $E_{th}$ and distances $d$ are listed in Table \ref{tab:counts_chooz}. The signal is integrated up to an energy of 5\,keV where the contribution to the signal is negligible. Count rates are given per day and kg as well as per day and detector array (CaWO$_4$: 6.84\,g, Al$_2$O$_3$: 4.41\,g).  A signal rate of up to $\sim10$\,counts per array and day is expected for CaWO$_4$ target calorimeters. 

Due to the relatively high rates predicted at such sites, the detection of CNNS with a small-scale detector of low threshold ($\sim$10\,eV) at a moderate distance from the core is clearly in reach.

\subsubsection{Background level}\label{sec:case1_background}

We consider a shallow experimental site with a small overburden to shield against cosmogenic backgrounds at $d\approx15-100\,$m from the reactor core. Possible candidate sites are, e.g. a room in the basement of a building outside the reactor containment, an artificial overburden outside the reactor building or even a site outside the reactor plant. At such places, the reactor-correlated gamma and neutron backgrounds are considered as negligible due to the large distance and  significant shieldings. In the following we concentrate on uncorrelated backgrounds which at shallow sites are dominated by muon-induced events  \cite{Gastrich:2015owx}. Plenty of experimental data describing detectors operated in shallow or  above-ground low-background environments exist in the literature, mostly for Ge detectors. The following total background levels are reached in selected experiments:  0.4\,counts/[kg keV day] at the shallow underground lab in Heidelberg  \cite{Heusser:2015ifa},  5\,counts/[kg keV day] at the Dortmund low-background facility \cite{Gastrich:2015owx}, $\sim20$\,counts/[kg keV day] at the ARC in Seibersdorf \cite{Schwaiger2002375} and 200\,counts/[kg keV day] in the CDMS experiment operated at the Stanford underground facility \cite{Akerib:2010pv} (black dotted lines). All values correspond to the rates in the lowest energy bin of the respective experiment.  The grey band in Fig. \ref{fig:spectrum_chooz} indicates the uncertainty in the observed background level depending on the individual site, the overburden and the  shielding design. The lowest energy threshold ($\sim500\,$eV) among the listed experiments is achieved by CDMS \cite{Akerib:2010pv}.

We use the highest background level reported as a conservative upper limit for the sensitivity studies. Even more conservative, we do not consider the additional background-rejection capability of the inner and outer cryogenic veto. As shown in  chapter \ref{sec:innerVeto_holder} and \ref{sec:outer_veto} by a dedicated MC study, the cryogenic fiducial-volume detector reduces surface, gamma and neutron  backgrounds by factors of $\gtrsim10^3$ and $\sim10$, respectively, in the target volume. In the following, the (flat) background rate of 200\,counts/[kg keV day]  is referred to as the benchmark. 

In case of  CaWO$_4$ the CNNS signal is  ${2-3}$ orders of magnitude above the conservative benchmark background whereas in case of  Al$_2$O$_3$ the signal-to-background ratio is much smaller (factor of 1-5), see Fig. \ref{fig:spectrum_chooz}. The multi-target approach, therefore, is a powerful tool to actively discriminate neutrino-induced signals from backgrounds.  In particular, it allows to identify possible ultimate exponentially shaped, signal-like backgrounds. 

\subsubsection{Experimental site and discovery potential}
An extensive likelihood study is performed to investigate the discovery potential of CNNS with the proposed small-scale experiment. We consider one  CaWO$_4$  (total mass: 6.84\,g) and one Al$_2$O$_3$ (total mass: 4.41\,g) calorimeter array inside the inner and outer active cryogenic veto (see Fig.  \ref{fig:outerveto}).  The benchmark background level is assumed and, conservatively, the rejection capability of the surface veto is not used for the background estimation. Three different thresholds are studied (5, 10 and 20\,eV) which, however, have only a minor impact on the discovery potential.  We define three scenarios: 
\begin{itemize} 
\item \textbf{Near case:} A distance of 15\,m from the reactor core - a  site within the reactor containment. Highest count rates are expected, but there are tough requirements for the shielding against correlated backgrounds. The access is restricted and strict safety regulations have to be considered. 
\item \textbf{Medium case:} A distance of 40\,m from the reactor core - outside the containment and the reactor building. Possibly a shallow site in a adjoining building or a dedicated site with an artificial overburden. Easier access and a better infrastructure.
 \item \textbf{Far case:} A distance of 100\,m from the reactor core - far away from the critical reactor components, possibly outside the entire power-plant area. Straightforward access and plenty of possible sites.  
\end{itemize}

\begin{figure}
\centering
\includegraphics[width=0.5\textwidth]{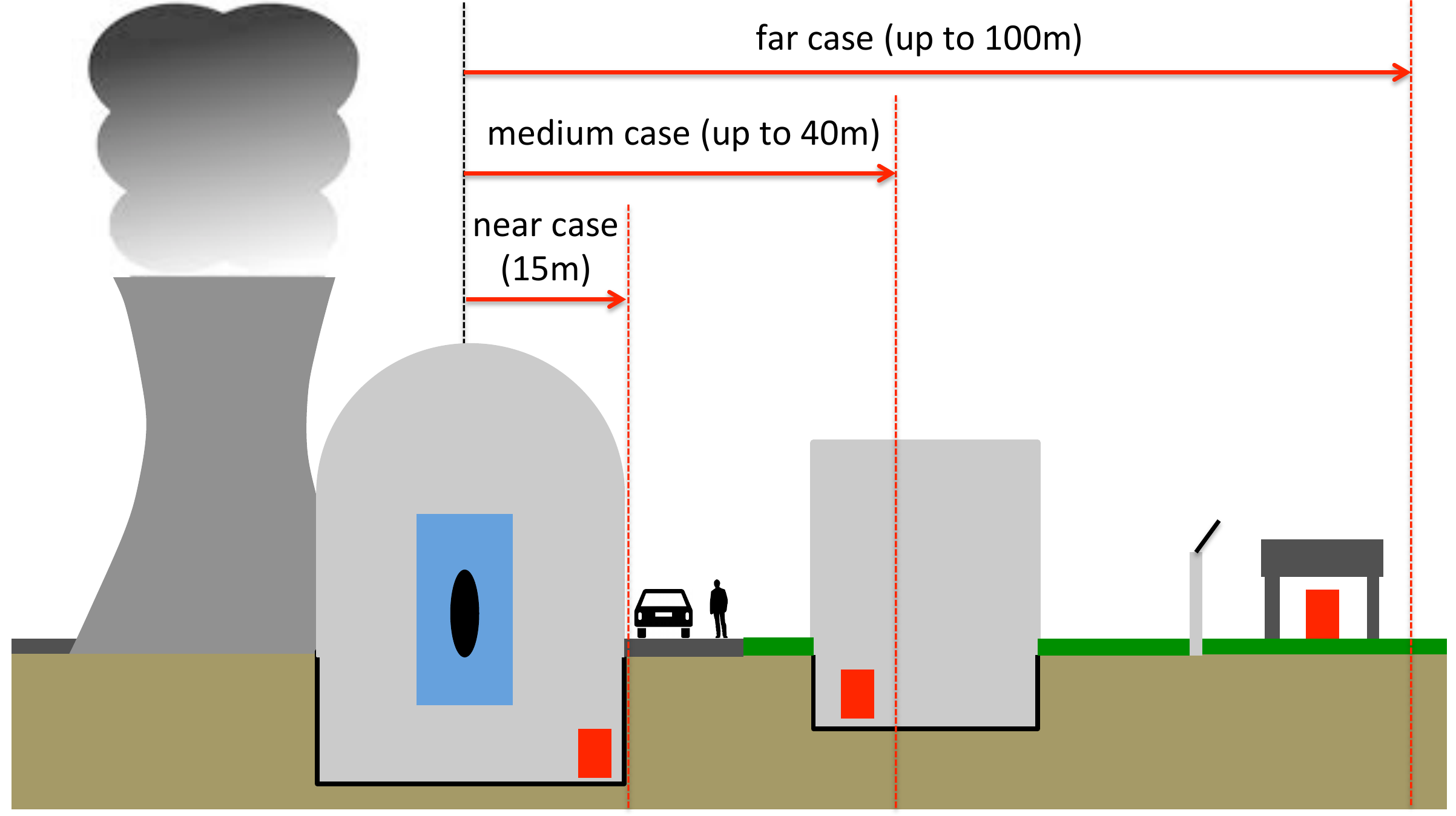}
\caption{Artist view of a typical nuclear power plant with possible experimental sites (red boxes) for the 3 different scenarios (see text).  }
\label{fig:spectra}
\end{figure}

\begin{figure}
\centering
\includegraphics[width=0.5\textwidth]{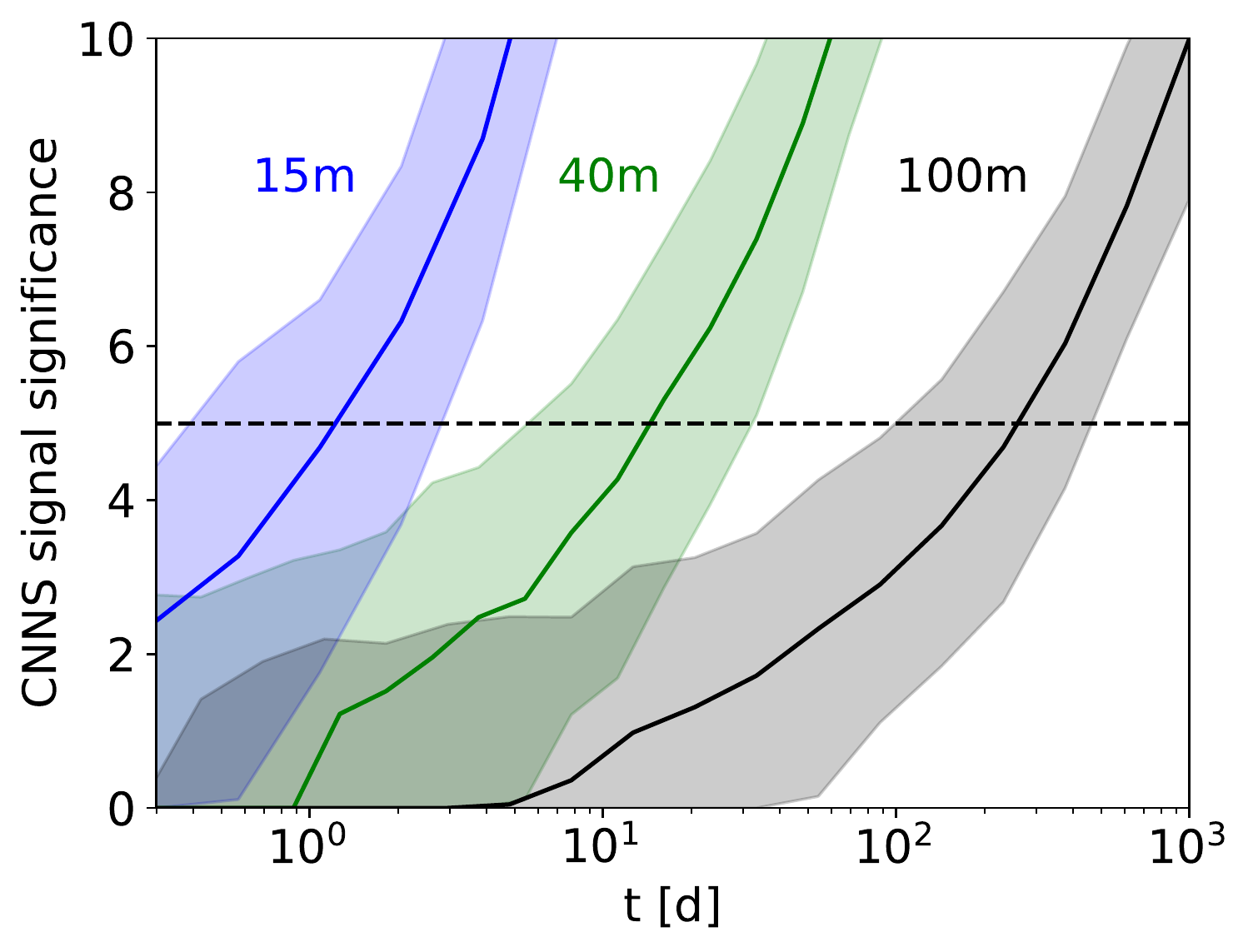}
\caption{Discovery potential of CNNS vs. time at a 4~GW reactor core from the likelihood ratio analysis described in the text. The combination of one CaWO$_4$ and one Al$_2$O$_3$ calorimeter array is investigated assuming the benchmark background level of $200$~counts/[kg keV day]. The full lines indicate the median discovery probability for a nominal energy threshold of 10\,eV, the bands show the 90\,\% confidence intervals. Three cases for the reactor distance $d$ are considered (see text): near case (blue), medium case (green) and far case (grey).  Varying the threshold to 5 and 20\,eV, respectively, has only a minor impact on the discovery potential (see text).}
\label{fig:likelihood_chooz}
\end{figure}

For each case, spectra are randomly generated for a large number of varying exposures. The results of this MC simulation are studied with a likelihood ratio analysis. In every MC experiment, one spectrum each is generated for the CaWO$_4$ and Al$_2$O$_3$ arrays. The unbinned likelihood of a model's parameters is calculated as a product over the individual likelihoods for each event in both spectra and the Poisson likelihood for observing this total event numer (Extended Maximum Likelihood method). The single event likelihood is proportional to the sum of the signal and background rates for the given parameter values. 
Two very simple models are considered: the free model has two parameters, namely the level of the flat background and the strength of the CNNS signal relative to the standard model expectation. In the null model, the CNNS signal strength is held at zero.
The maximum likelihood of each model at the best fit parameter values is denoted $\mathcal{L}_\mathrm{free}$ and $\mathcal{L}_\mathrm{null}$ respectively. 
Since the two models are nested with one additional parameter in the free model, the likelihood ratio test statistic 
\[W = 2\log\frac{\mathcal{L}_\mathrm{free}}{\mathcal{L}_\mathrm{null}}\]
follows a $\chi^2$-distribution with one degree of freedom (by Wilks' theorem). The square root of the test statistic therefore follows a standard normal distribution, so that the statistical significance in $\sigma$ of the claim of a CNNS signal with nonzero cross-section in addition to the assumed flat background is directly given by $\sqrt{W}$ for each pair of spectra.

Fig. \ref{fig:likelihood_chooz} shows the resulting discovery potential of the 3 scenarios. The full lines indicate the median discovery potential as derived from the MC data, using an energy threshold of 10\,eV. The bands give the 90\,\% confidence intervals.   All three scenarios show a very promising potential for the discovery (5$\sigma$) of CNNS - in the near case within $\sim1$\,day, in the medium case within $\lesssim2$\,weeks and in the far case within $\sim1$\,year of measuring time.  Improving the threshold to 5\,eV reduces the measuring time necessary for a 5$\sigma$ discovery by a factor of $\sim1.3$, in average for the three scenarios discussed. For a threshold of 20\,eV,  $\sim1.6$ times longer measurements are required.

Systematic deviations of background and signal rates have only minor influence on the discovery potential:
Repeating the simulations with 20\,\% higher and lower background level yield 10\,\% higher and lower times to discovery, respectively. A 5\,\% stronger signal makes discovery faster by 5\,\%, while a 5\,\% weaker signal requires 7.5\,\% more measuring time. 

To study the impact of a non-flat background on the discovery potential we use data from a CDMS detector operated at a shallow laboratory \cite{Akerib:2010pv}. The measured spectrum was fitted with an exponential below 10\,keV and extrapolated exponentially to beyond the energy threshold of 500\,eV. The fitted background model corresponds to an exponential component rising above a flat background of 2.5/\,keV\,kg\,d at around 2\,keV and reaching 700/\,keV\,kg\,d at zero energy. Using this background level in the likelihood study has only a minor impact. The measuring time required for a 5$\sigma$ discovery in the three scenarios increase moderately by a factor of $\sim2.5$.

Background studies including dedicated measurements on the individual sites and detailed MC simulations are required to find the most suitable site. At the medium and far sites, expected backgrounds are rather straightforward, while for the near site a proper understanding of the possibly remaining reactor-correlated backgrounds is needed. The near site, however, would - despite a rapid discovery of CNNS - allow a precision measurement (statistical error on a percent level) of the cross-section predicted by the Standard Model within a measuring time of one year. Impressively, this can be performed by a detector with a total target mass of $\sim10$\,g, given the necessary control of systematics.

\subsection{Case 2: At a thermal research reactor}
\begin{figure}
\centering
\includegraphics[width=0.5\textwidth]{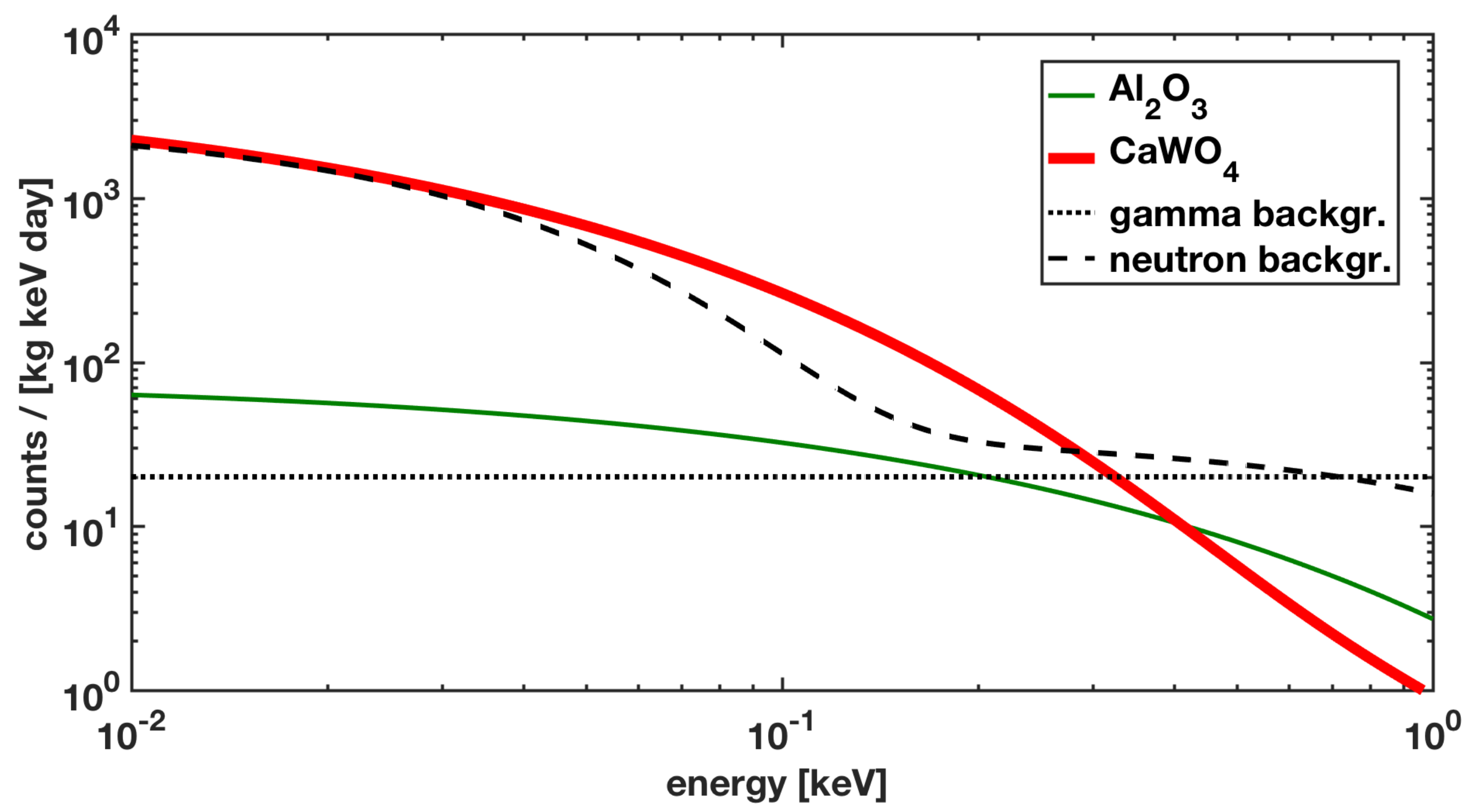}
\caption{Expected CNNS count rates at the FRM2 research reactor with a thermal power of 20\,MW at a distance 3\,m. The isotopic fractions of the neutrino emission is adjusted to 96\% from $^{235}$U and 4\% from $^{238}$U to account for the different fuel composition at FRM2. The thick red and green lines show the CNNS rates on CaWO$_4$ and Al$_2$O$_3$, respectively. In black, the simulated research-reactor background spectra for neutrons (scaled to 10\%, see text) and gammas from~\cite{2016arXiv160902066M} are shown.}
\label{fig:frm2_rates}
\end{figure}

To study the possibility of detecting CNNS at a thermal research reactor, both the signal and background spectrum were adapted to the altered conditions. The signal expectation was calculated for the fuel composition found at FRM2 (96\% of $^{235}$U, 0\% of $^{239}$Pu and 4\% of $^{238}$U) \cite{2012RScI...83g5109W} which does not change the signal shape appreciably. Our fiducial model is a 20~MW reactor at a distance of 3~m to the detector, which corresponds to a neutrino flux reduced by $\sim~2.4$ with respect to the medium distance case at the power reactor. The close proximity to the research reactor implies great uncertainty regarding the composition and shape of backgrounds correlated to the reactor power, and thus to the neutrino signal. We use the results of a  detailed background measurement and simulation campaign~\cite{2016arXiv160902066M} by the MINER collaboration for our background estimate. In the framework of the MINER experiment it is planned to deploy an array of Ge cryogenic detectors in close proximity to a 1\,MW research reactor \cite{2016arXiv160902066M}. Between the reactor core and the detectors, several layers of shielding made of graphite, borated high-density polyethylene, Pb and Cu are placed. These conditions are assumed to be comparable to the setup we are investigating in this context.
Fig. \ref{fig:frm2_rates} shows the expected CNNS rates at the research reactor at a distance of  3\,m from the core, along with the neutron (scaled to 10\%, see below) and gamma backgrounds from~\cite{2016arXiv160902066M}.  Compared to those, cosmogenic backgrounds and environmental radioactivity appear to be  sub-dominant in shallow labs (compare to section \ref{sec:case1_background}) and are neglected in this basic feasibility study. 

The similarity of the signal shape to the reactor correlated neutron background makes the detection of CNNS challenging in this environment even with extensive reactor ON/OFF measurements. Under these conditions, a multi target approach can be beneficial because of the material dependence of CNNS (through the cross-section $\propto N^2$). In both CaWO$_4$ and Al$_2$O$_3$, neutrons are expected to scatter predominantly off the light oxygen nuclei, leaving a comparable signature. On the other hand, the Al$_2$O$_3$ array does not contribute meaningfully to the CNNS signal measurement, but yields an important measurement of the background rate as a function of energy. This information helps to break the degeneracy between the CNNS signal and the neutron background, which are very similar in shape.

\begin{figure}
\centering
\includegraphics[width=0.45\textwidth]{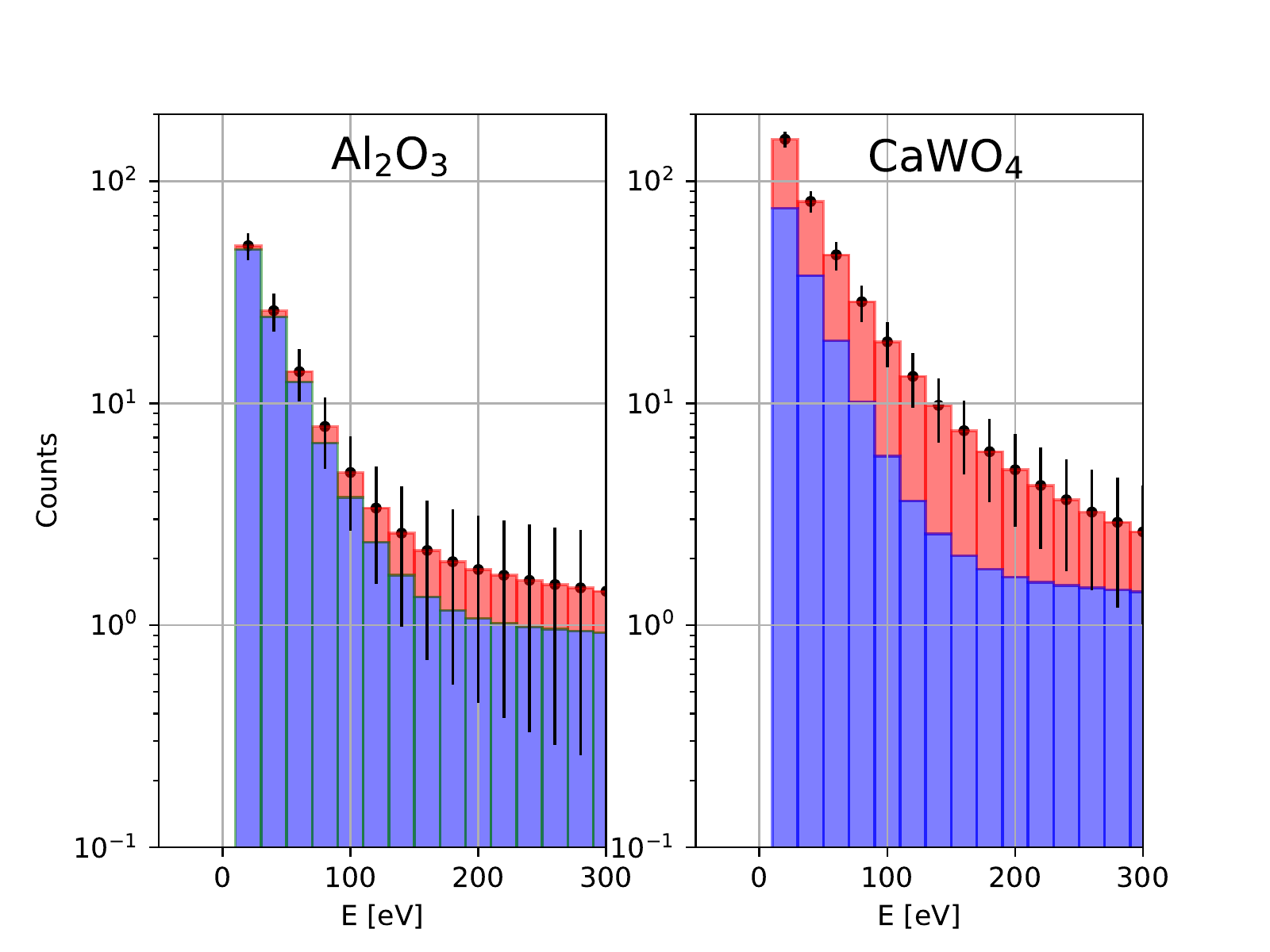}
\caption{Expected spectra at a thermal research reactor after one year of measurement in the Al$_2$O$_3$ array (left) and in the CaWO$_4$ array (right). Shown are the expected background counts (blue) from the MINER background scaled to 10\%, and the expected CNNS signal counts (in red). The error bars show the expected  fluctuations. The background spectra are assumed to be identical in the two materials, scaled only by the respective exposure (higher in CaWO$_4$ due to the higher density of the material). The CNNS signal is strongly enhanced in CaWO$_4$ due to the neutron-rich W nucleus. The Al$_2$O$_3$ array with a similar neutron response can serve for an in-situ background characterization.}
\label{fig:frm2_spectra}
\end{figure}

Fig. \ref{fig:frm2_spectra} shows the expected counts in the Al$_2$O$_3$ array (left) and in the CaWO$_4$ array (right) after one year of measurement in the described conditions. The similarity of the signal and background shapes in CaWO$_4$ is apparent. The signal in Al$_2$O$_3$ is overwhelmed by the Poisson fluctuations, so the Al$_2$O$_3$ array only contributes to the determination of the background level.

\begin{figure}
\centering
\includegraphics[width=0.4\textwidth]{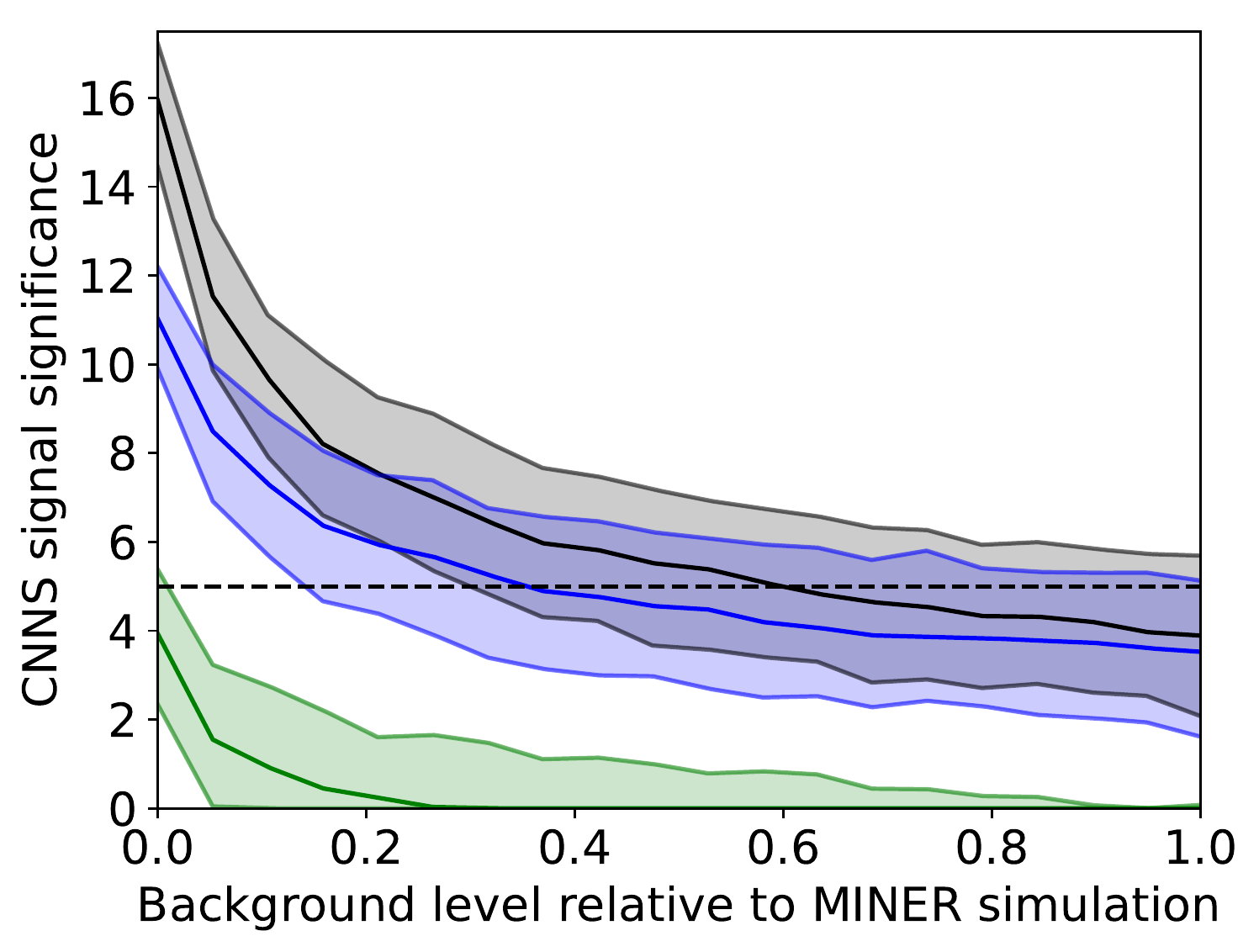}
\caption{CNNS discovery potential at a research reactor vs. background strength. The background is based on a detailed neutron simulation performed within the MINER collaboration (Fig. 14 in~\cite{2016arXiv160902066M}). The detection significance is plotted separately for the Al$_2$O$_3$ (green) and CaWO$_4$ (blue) arrays. The combination of both (black) considerably enhances the detection significance. Full lines represent the median discovery potential, the bands constraint by  thin lines are 90\,\% confidence intervals as derived by the MC simulation. }
\label{fig:frm2_likelihood}
\end{figure}

For the likelihood study, we assume a fixed live time of one year with the CaWO$_4$ and Al$_2$O$_3$ arrays and show the detection significance (computed as above) as a function of the background level instead. The simulated background spectrum is a scaled version of the MINER neutron background. The shape of the background spectrum is assumed to be known for the likelihood models. To show the added benefit of the Al$_2$O$_3$ array for background characterization, we plot the significance obtained by each detector material separately (green, blue) along with the combined significance (obtained as above, black) in Fig. \ref{fig:frm2_likelihood}. Full lines indicate the median discovery potential, the bands are 90\,\% confidence intervals. The background level has to be reduced significantly with respect to the level reported in~\cite{2016arXiv160902066M} to allow a $5\sigma$-detection within one year. The ``background-only'' information provided by the Al$_2$O$_3$ array considerably relaxes the background requirements, so that a detection after one year becomes very likely with a background below $\sim 30\%$ of the MINER neutron background, and feasible with a background below $\sim 60\%$.

\subsection{Case 3: Using a neutrino source}

\begin{figure}
\centering
\includegraphics[width=0.5\textwidth]{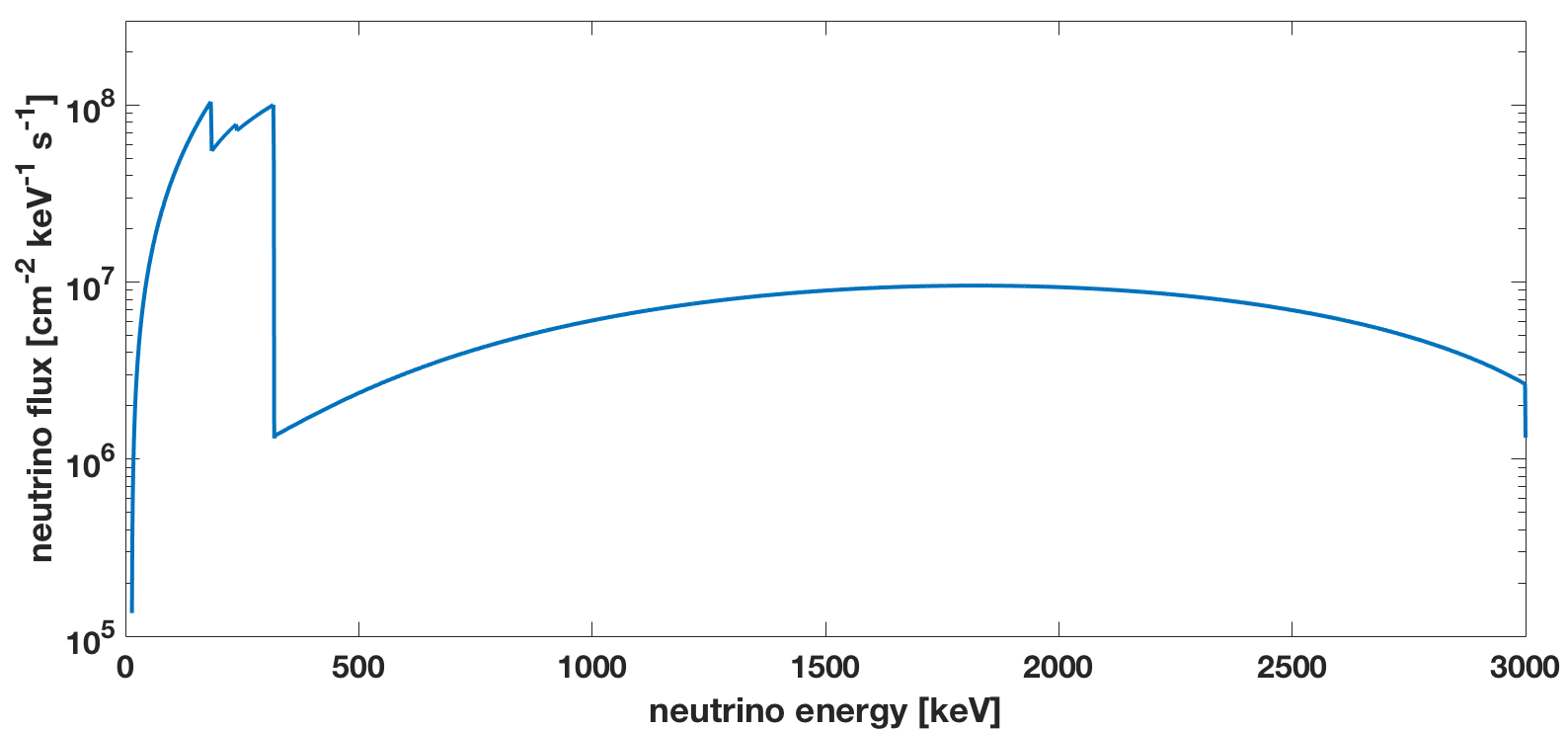}
\caption{Neutrino flux from the Ce neutrino source with an assumed activity of 150 kCi. The low-energy spectrum up to $\sim 300$\,keV corresponds to the initial $^{144}$Ce decay, with the high-energy spectrum up to $Q=3$\,MeV stems from the decay of the daughter nucleus $^{144}$Pr.}
\label{fig:Ce_flux}
\end{figure}

\begin{figure}
\centering
\includegraphics[width=0.5\textwidth]{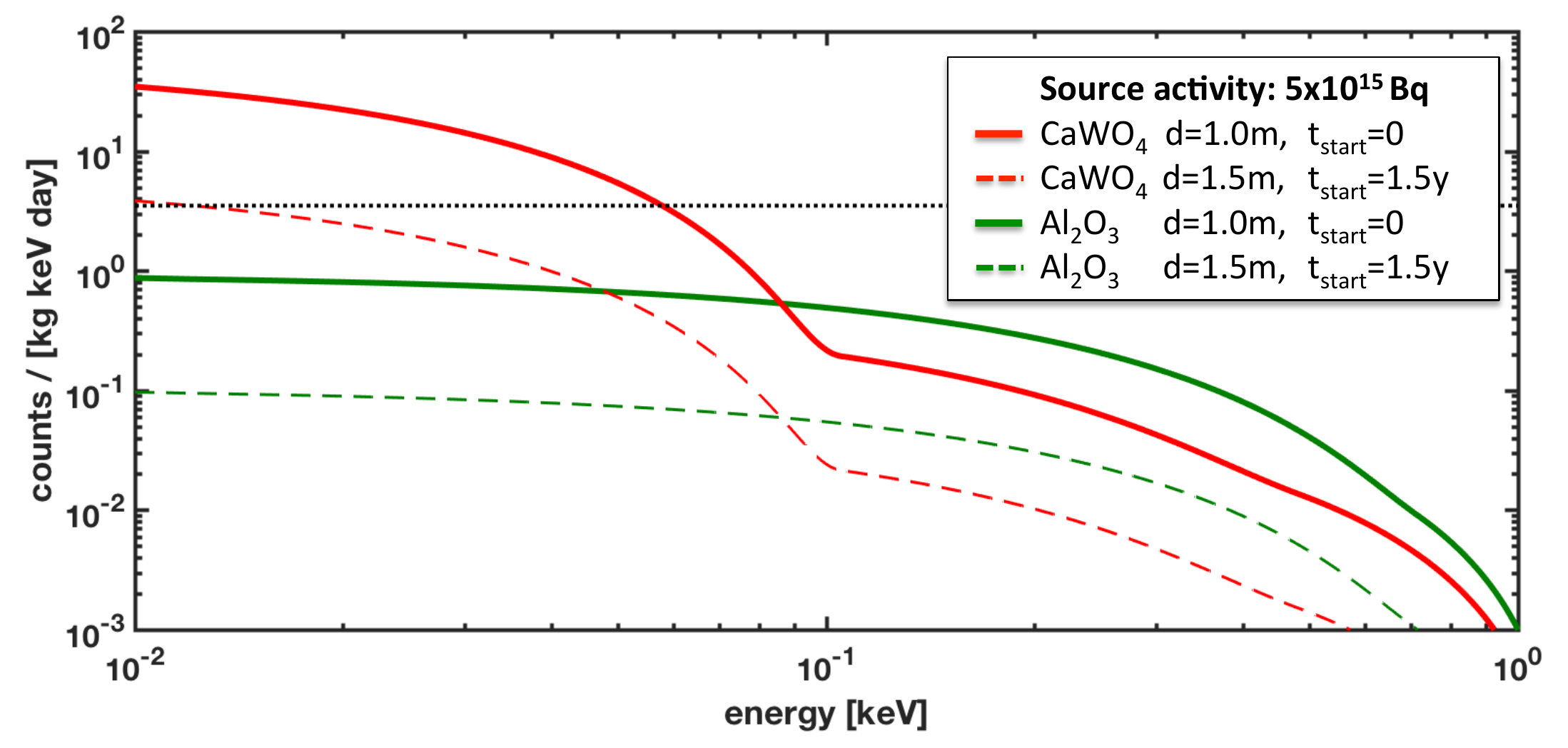}
\caption{Count rates expected from a benchmark $^{144}$Ce neutrino source with an initial activity ($t_{start}=0$) of $5\cdot10^{15}$\,Bq. The dotted lines indicate the worst-case scenario with a distance of 1.5\,m and a measurement after two half-lifes of the source ($t_{start}=1.5$\,y). The black dotted line indicates the background level achieved with CaWO$_4$ in the CRESST setup at LNGS, Italy \cite{strauss:2014part2}. }
\label{fig:Ce_rates}
\end{figure}

The detection of CNNS using a radioactive neutrino source is a scenario which poses quite different challenges compared to the other considered cases. To evaluate the new situation, we assume a neutrino source similar to the source proposed for the SOX experiment~\cite{1742-6596-718-6-062066}, specifically we show in  Fig. \ref{fig:Ce_flux} the neutrino spectrum of a $^{144}$Ce source with an initial activity of 150\,kCi. The low-energy neutrinos ($\lesssim 300$~keV) originate in the initial decay of $^{144}$Ce with a half life of 285 days, while the broad neutrino spectrum up to a Q value about 3\,MeV originates in the fast decay of the daughter nucleus $^{144}$Pr. 
The low-energy neutrinos do not produce a detectable W recoil in the detectors considered here. With a 10\,eV threshold, a CaWO$_4$ detector is sensitive mostly to neutrinos above 1\,MeV.
Therefore, even with an optimistic shielding scenario (source distance of 1\,m) to stop residual gammas and neutrons from source impurities, the recoil rates are more than an order of magnitude below the far case of the power reactor scenario, as shown in Fig. \ref{fig:Ce_rates}. 

The background level shown there is the measured intrinsic background of CRESST crystals, which constitutes the best reasonably achievable background level. We point out that any  parasitic, radioactive contamination of the neutrino source which may produce additional neutron and gamma background is neglected in this study. With this optimistic assumption, detection of CNNS becomes feasible, but still suffers from low count rates.

\begin{figure}
\centering
\includegraphics[width=0.4\textwidth]{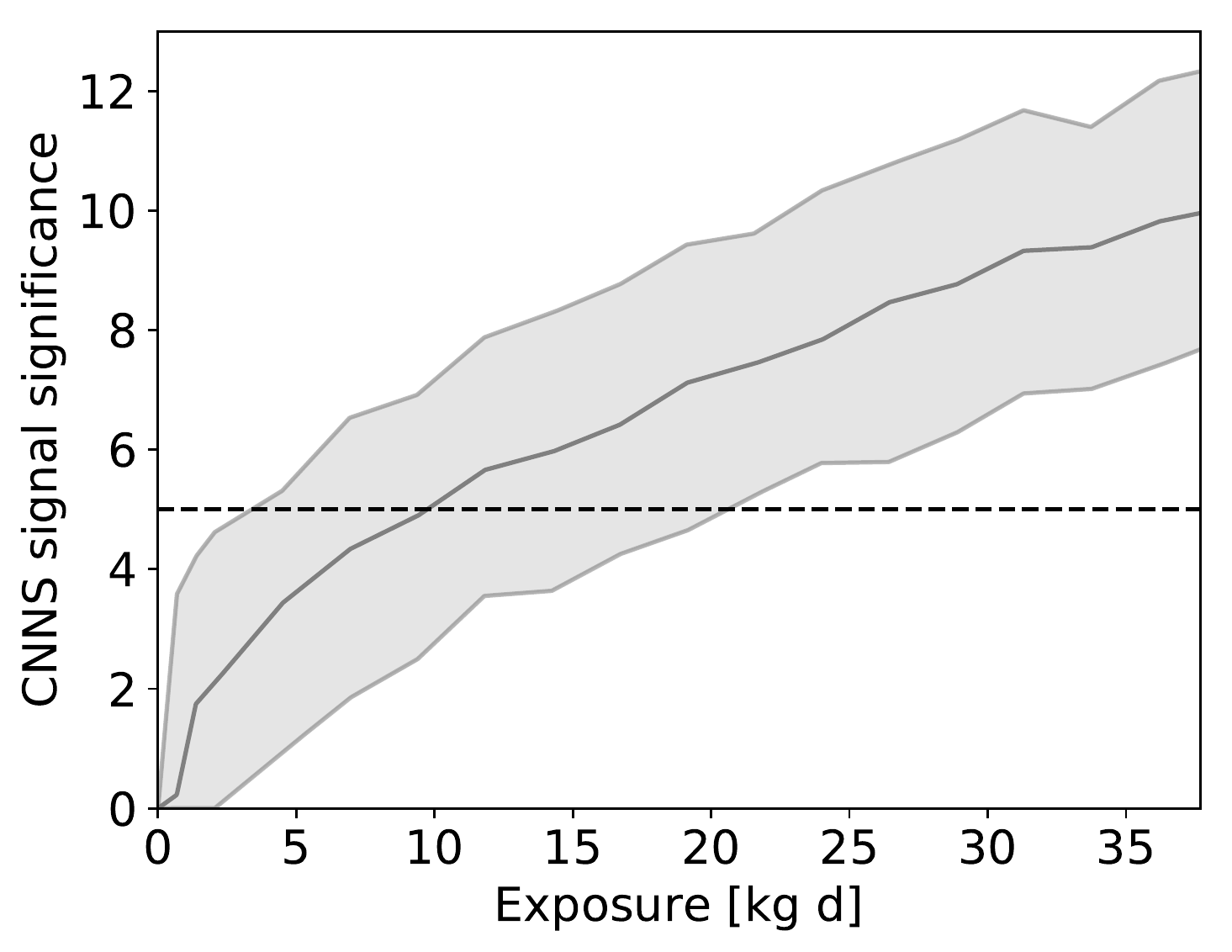}
\caption{Discovery potential of CNNS  vs. exposure using a neutrino source, derived by a dedicated likelihood analysis. The full line represents the median discovery potential, the band is the 90\,\% confidence interval. An exposure of $\sim$10\,kg\,d is required for a detection of CNNS.}
\label{fig:Ce_likelihood}
\end{figure}

Fig. \ref{fig:Ce_likelihood} shows the discovery potential of CNNS at 1\,m from the benchmark radioactive neutrino source as a function of the exposure collected at full activity (150\,kCi), obtained with the likelihood ratio method described above. Detection comes in reach with an exposure of $\sim 10$\,kg\,d. Such exposures are feasible, however require a larger detector mass. The exposure has to be collected within a few half-lives of the source isotope ($t_{1/2} = 285$\,d), which necessitates a larger array, e.g. ${10\times10}$ cubes (${\sim50}$\,g).

\section{Summary and Outlook}
The smallness of gram-scale calorimeters offers the following significant advantages:  1)~very low
energy thresholds down to the 10\,eV-regime and presumably below, 2)~ the possibilty of an encapsulation of the small
calorimeters by other cryogenic devices which act as  anti-coincidence vetos
 and 3)~ the ability to operate the detectors at the surface in a relatively 
high-rate environment. These advantages are demonstrated experimentally by a measurement with a prototype 0.5\,g Al$_2$O$_3$ calorimeter which reaches a threshold of $\sim20$\,eV.  Based on that, we propose a new detector concept here:  a gram-scale fiducial-volume cryogenic detector for the detection of CNNS, called the {\large$\nu$}-cleus experiment.

A basic version of the detector, consisting of two 3x3 calorimeter arrays made of CaWO$_4$ and Al$_2$O$_3$ crystals with a total mass of $\sim10$\,g, has a high potential for a rapid discovery of CNNS. We study various experimental scenarios for this new technology: an installation at a nuclear power plant, at a thermal research reactor and close to a radioactive neutrino source. We conclude that all three methods allow a detection of  CNNS, however the first scenario clearly shows the highest potential.

We investigate the operation of the 10\,g detector at a distance of $\sim$40\,m from a nuclear power reactor with a  thermal power of 4\,GW. This corresponds to an experimental site outside the reactor containment, and is therefore rather straight-forward in terms of background levels, infrastructure and access. As shown by a dedicated likelihood analysis, the rate is still sufficiently high to achieve a 5$\sigma$ discovery within a measuring time of $\lesssim2$\,weeks. 

The detector placed at a well-shielded site within the reactor containment, e.g. at a distance of $\sim$15\,m from the core, would give the unique possibility for precision measurements of the  CNNS cross-section and probe, e.g. the Weinberg angle at low momentum transfers ~\cite{Lindner:2016wff}. Furthermore, since a discovery of CNNS is possible within a day, this technology can be used for real-time monitoring of nuclear reactors. Such a small-scale experimental setup could provide a comprehensive surveillance system for non-proliferation and accident prevention at the $\sim500$ nuclear reactors world-wide. 

The detector is designed such to be scaled up in a relatively simple way due to the use of production techniques of the semiconductor industry. We point out that total target masses of $\mathcal{O}$(1\,kg) are feasible with the design principle given here. This enables new approaches in rare events searches, such as the search for MeV-scale dark matter and flavour-independent precision measurements of the solar neutrino flux. Operating such a kg-scale detector with energy thresholds in the 10\,eV regime at a power reactor  opens the door to a variety of new physics beyond the Standard Model of Particle physics.

\bibliography{MeVpaper}

\begin{thebibliography}{10}
\providecommand{\url}[1]{{#1}}
\providecommand{\urlprefix}{URL }
\expandafter\ifx\csname urlstyle\endcsname\relax
  \providecommand{\doi}[1]{DOI \discretionary{}{}{}#1}\else
  \providecommand{\doi}{DOI \discretionary{}{}{}\begingroup
  \urlstyle{rm}\Url}\fi

\bibitem{PhysRevD.9.1389}
D.Z. Freedman, Phys. Rev. D \textbf{9}, 1389 (1974)

\bibitem{PhysRevD.30.2295}
A.~Drukier, L.~Stodolsky, Phys. Rev. D \textbf{30}, 2295 (1984)

\bibitem{Akimov:2015nza}
D.~Akimov, et~al.,   (2015).
\newblock \urlprefix\url{https://arxiv.org/abs/1509.08702}

\bibitem{Aguilar-Arevalo:2016khx}
A.~Aguilar-Arevalo, et~al., J. Phys. Conf. Ser. \textbf{761}(1), 012057 (2016).
\newblock \doi{10.1088/1742-6596/761/1/012057}

\bibitem{Moroni:2014wia}
G.~Fernandez~Moroni, J.~Estrada, E.E. Paolini, G.~Cancelo, J.~Tiffenberg,
  J.~Molina, Phys. Rev. \textbf{D91}(7), 072001 (2015).
\newblock \doi{10.1103/PhysRevD.91.072001}

\bibitem{Kerman:2016jqp}
S.~Kerman, V.~Sharma, M.~Deniz, H.T. Wong, J.W. Chen, H.B. Li, S.T. Lin, C.P.
  Liu, Q.~Yue, Phys. Rev. \textbf{D93}(11), 113006 (2016).
\newblock \doi{10.1103/PhysRevD.93.113006}

\bibitem{2016arXiv160902066M}
G.~{Agnolet}, et~al., ArXiv e-prints  (2016).
\newblock \urlprefix\url{http://adsabs.harvard.edu/abs/2016arXiv160902066M}

\bibitem{Billard:2016giu}
J.~Billard, et~al.,   (2016).
\newblock \urlprefix\url{https://arxiv.org/abs/1612.09035}

\bibitem{cnns_letter}
R.~Strauss, et~al., Phys. Rev. D \textbf{96}, 022009 (2017).
\newblock \urlprefix\url{{https://inspirehep.net/record/1591648}}

\bibitem{Lindner:2016wff}
M.~Lindner, et~al., JHEP \textbf{2017}(3), 979 (2017)

\bibitem{2015arXiv151103934B}
N.~{Berger}, et~al., ArXiv e-prints  (2015).
\newblock \urlprefix\url{http://adsabs.harvard.edu/abs/2015arXiv151103934B}

\bibitem{PhysRevD.72.073003}
J.~Erler, M.J. Ramsey-Musolf, Phys. Rev. D \textbf{72}, 073003 (2005)

\bibitem{0034-4885-76-4-044201}
T.~Ohlsson, Reports on Progress in Physics \textbf{76}(4), 044201 (2013)

\bibitem{santamaria2003present}
A.~Santamaria, C.~Pe{\~n}a~Garay, S.~Davidson, N.~Rius~Dionis, Journal of High
  Energy Physics, 2003, vol. 0303, p. 011  (2003)

\bibitem{PhysRevD.86.013004}
A.J. Anderson, et~al., Phys. Rev. D \textbf{86}, 013004 (2012)

\bibitem{doi:10.1142/S0217732305017482}
H.T. Wong, et~al., Modern Physics Letters A \textbf{20}(15), 1103 (2005)

\bibitem{rothe_master}
J.~Rothe, Master's thesis, LMU M\"unchen, https://publications.mppmu.mpg.de
  (MPP-2016-391) (2016)

\bibitem{Angloher:2015ewa}
G.~Angloher, et~al., Eur. Phys. J. \textbf{C76}(1), 25 (2016)

\bibitem{moseley_xray}
S.H. Moseley, et~al., J. Appl. Phys. \textbf{56}(5), 1257 (1984)

\bibitem{Pyle:2015pya}
M.~{Pyle}, E.~{Figueroa-Feliciano}, B.~{Sadoulet} (2015).
\newblock \urlprefix\url{http://adsabs.harvard.edu/abs/2015arXiv150301200P}

\bibitem{Probst:1995fk}
F.~Pr{\"o}bst, et~al., J. Low Temp. Phys. \textbf{100}(1-2), 69 (1995)

\bibitem{Angloher:2014myn}
G.~Angloher, et~al., Eur. Phys. J. \textbf{C74}(12), 3184 (2014)

\bibitem{Meier2000350}
O.~Meier, et~al., Nucl. Instrum. Meth. \textbf{A444}, 350 (2000)

\bibitem{Angloher200243}
G.~Angloher, et~al., Astropart. Phys. \textbf{18}(1), 43  (2002)

\bibitem{Strauss2016}
R.~Strauss, et~al., Nucl. Instrum. Meth. \textbf{A845}, 414 (2017)

\bibitem{Angloher2016}
G.~Angloher, et~al., J. Low Temp. Phys. \textbf{184}(1), 323 (2016)

\bibitem{Angloher:2012vn}
G.~Angloher, et~al., Eur. Phys. J. \textbf{C72}, 1971 (2012)

\bibitem{Agnese:2016cpb}
R.~Agnese, et~al., Phys. Rev. D \textbf{95}, 082002 (2017)

\bibitem{Strauss:2014ab}
R.~Strauss, et~al., Eur.Phys.J. \textbf{C74}(7), 2957 (2014)

\bibitem{g4a}
S.~Agostinelli, et~al., Nucl. Instrum. Meth. A \textbf{506}(3), 250  (2003).
\newblock \doi{10.1016/S0168-9002(03)01368-8}

\bibitem{g4b}
J.~Allison, et~al., IEEE Transactions on Nuclear Science \textbf{53}(1), 270
  (2006).
\newblock \doi{10.1109/TNS.2006.869826}

\bibitem{g4c}
{Geant4 Low Energy Electromagnetic Physics Working Group}.
\newblock Low energy physics lists.
\newblock \url{https://twiki.cern.ch/twiki/bin/view/Geant4/LowePhysicsLists}
  (2016)

\bibitem{strauss:2014part2}
R.~Strauss, et~al., JCAP \textbf{1506}(06), 030 (2015)

\bibitem{firestone}
R.B. Firestone, \emph{Table of Isotopes}, 8th edn. (Wiley, 1996)

\bibitem{Strauss2014}
R.~Strauss, et~al., Eur. Phys. J. C \textbf{74}(7), 2957 (2014).
\newblock \doi{10.1140/epjc/s10052-014-2957-5}

\bibitem{Gastrich:2015owx}
H.~Gastrich, et~al., Appl. Radiat. Isot. \textbf{112}, 165 (2016)

\bibitem{2012RScI...83g3113D}
M.A. {Dobbs}, et~al., Review of Scientific Instruments \textbf{83}(7), 073113
  (2012)

\bibitem{Kopeikin:2004cn}
V.~Kopeikin, L.~Mikaelyan, V.~Sinev, Phys. Atom. Nucl. \textbf{67}, 1892
  (2004).
\newblock [Yad. Fiz.67,1916(2004)]

\bibitem{1989NuPhA.503..136T}
O.~{Tengblad}, et~al., Nuclear Physics A \textbf{503}, 136 (1989)

\bibitem{Akerib:2010pv}
D.S. Akerib, et~al., Phys. Rev. \textbf{D82}, 122004 (2010)

\bibitem{Schwaiger2002375}
M.~Schwaiger, F.~Steger, T.~Schroettner, C.~Schmitzer, Applied Radiation and
  Isotopes \textbf{56}(1{\^a}2), 375  (2002)

\bibitem{Heusser:2015ifa}
G.~Heusser, et~al., Eur. Phys. J. \textbf{C75}(11), 531 (2015)

\bibitem{2012RScI...83g5109W}
J.~{Wuttke}, et~al., Review of Scientific Instruments \textbf{83}(7), 075109
  (2012).
\newblock \doi{10.1063/1.4732806}

\bibitem{1742-6596-718-6-062066}
M.~Vivier, et~al., Journal of Physics: Conference Series \textbf{718}(6),
  062066 (2016)

\end{thebibliography}

\newpage

\end{document}